\documentclass[a4paper,fleqn,usenatbib,useAMS]{mnras}

\usepackage{graphicx,multicol}
\usepackage{times}

\usepackage{amsmath} 
\usepackage{amssymb}   

\let\baraccent=\= 
\renewcommand{\=}[1]{\stackrel{#1}{=}} 

\title[Externally driven models]{Externally-driven plasma models as candidates for pulsar radio emission}

\author[Rahaman et al.]{
Sk. Minhajur Rahaman$^{1}$\thanks{E-mail: rahaman@ncra.tifr.res.in}, Dipanjan Mitra$^{1,2}$ , George I. Melikidze$^{2,3}$\\
$^{1}$ National Centre for Radio Astrophysics, Tata Institute of Fundamental Research, Post Bag 3, Ganeshkind,Pune-411007,India\\
$^{2}$ Janusz Gil Insitute of Astronomy, University of Zielona G\'ora, ul Szafrana 2, 65-516 Zielana G\'ora, Poland \\
$^{3}$ Evgeni Kharadze Georgian National Astrophysical Observatory, 0301, Abastumani, Georgia }

\date{Accepted XXX. Received YYY; in original form ZZZ}

\pubyear{2015}

\begin{document}
\label{firstpage}
\pagerange{\pageref{firstpage}--\pageref{lastpage}}
\maketitle

\begin{abstract}
Coherent radio emission from pulsars originates from excited plasma
waves in an ultra-relativistic and strongly magnetized
electron-positron pair plasma streaming along the open magnetic field
lines of the pulsar. Traditional coherent radio emission models have
relied on instabilities in this pair plasma. Recently alternative
models have been suggested. These models appeal to direct coupling of
the external electromagnetic field to the superluminal O-mode ($lt_2$
mode) during the time-dependent pair cascade process at the polar
gap. The objective of this work is to provide generic constraints on
plasma models based on $lt_2$ mode using realistic pulsar
parameters. We find that the very short timescale associated with pair
cascades does not allow $lt_{2}$ mode to be excited at radio
frequencies and the impulsive energy transfer can only increase the
kinetic spread (``temperature'') of the pair plasma
particles. Moreover, under homogeneous plasma conditions, plasma waves
on both branches of O-mode (i.e. superluminal $lt_2$ and subluminal
$lt_1$) cannot escape the plasma. In the strongly magnetized pair
plasma, only the extraordinary mode ($t$ mode) can escape freely. We
show that any generic fictitious mechanisms does not result in the
wave electric field of $t$ mode to have predominant orientation either parallel
or perpendicular to the magnetic field plane as observed. Such
fictitious mechanisms will inevitably lead to depolarization of
signals and cannot account for the highly polarized single pulses
observed in pulsars. We suggest coherent curvature radiation as a
promising candidate for pulsar radio emission mechanism.
\end{abstract}




\section{Introduction}
The coherent radio emission from pulsars is believed to originate from
an electron-positron pair plasma that streams relativistically outward
along the the open ambient magnetic field lines of the pulsar
magnetosphere. Traditional models of coherent radio emission appeal to
instabilities in this strongly magnetized and highly relativistic pair
plasma. Such instability driven mechanisms are broadly divided into
two classes viz., maser and antenna mechanisms
(\citealt{1975ARA&A..13..511G,1995JApA...16..137M}).  The former
assumes that radio emission is generated by certain types of plasma
instabilities that generate plasma waves capable of escaping the
magnetosphere, while the latter relies on coherent curvature radiation
as the main source of the observed radio waves. Recently a class of
externally driven models have been proposed that avoids the need for
plasma instability (\citealt{2020PhRvL.124x5101P} hereafter PTS20 ;
\citealt{2021MNRAS.500.4549M}; \citealt{2021arXiv210811702C}).

The preferred mechanism of pulsar radio emission should explain the
essential characteristics of the observed radio emission. Current
observations tightly constrains the location of the pulsar radio
emission to be far inside the magnetosphere, at least below 10\% of
the light cylinder
(\citealt{1991ApJ...370..643B,2017JApA...38...52M}). In this region
most of plasma instabilities are suppressed by the strong dipolar
ambient magnetic field of the pulsar. The only instability that can
develop is the two-stream instability which assumes wave-particle
interaction at the Cherenkov plasma resonance.  As the magnetic field
strength gets weaker with distance, nearer the light cylinder some
other instabilities like cyclotron and/or Cherenkov-curvature drift
instabilities can also develop
(e.g. \citealt{1991MNRAS.253..377K,1999MNRAS.305..338L}). These
instabilities are of the maser kind, and given the emission height
constraint they can be ruled out as possible candidates for pulsar
radio emission.

In addition to the location of the radio emission zone, polarisation
of the observed radiation provides an important constraint. The
observed radio waves are highly linearly polarised and the
polarisation vector is either parallel or perpendicular to the plane
in which the dipolar magnetic field lines lie
(\citealt{2014ApJ...794..105M} ; \citealt{2009ApJ...696L.141M}; \citealt{2001ApJ...549.1111L}).
Therefore, any pulsar radio-emission mechanism(s) should be based on
excitation of eigen modes of the strongly magnetised electron-positron
pair plasma which can propagate in the magnetosphere and then escape
from it by preserving the observed polarization features.

The eigen modes in strongly magnetized homogeneous pair plasma has
been extensively studied
(e.g. \citealt{1978ApJ...219..274H,1986ApJ...302..120A}). Here we
briefly summarize the behaviour of the eigen modes as discussed in
\cite{2003PhRvE..67b6407S}.  Generally, there are two eigen modes in
magnetised pair plasma, the so called ordinary (O-mode or $lt$ mode)
and extraordinary ($t$ mode) modes. The wave electric field (also
referred as polarisation vector) of O-mode lies in the plane of the
wave vector $\vec k$ and the ambient magnetic field $\vec B$ and has a
component along the ambient magnetic field. The $t$ mode always has a
purely transverse nature and its polarization vector is directed
perpendicular to $\vec k$ and $\vec B$ plane. The O-mode represent
longitudinal-transverse waves and have two branches, one subluminal
$lt_1$-mode and the other superluminal $lt_2$-mode. In the case of
strictly parallel propagation (with respect to the external magnetic
field) the superluminal branch coincides with the Langmuir wave and is
purely longitudinal. For larger angles the superluminal $lt_2$-mode
has longitudinal-transverse nature.

When propagating along the magnetic field, $lt_1$-mode is an
arbitrarily polarized purely transverse wave, whereas for oblique
propagation it becomes a longitudinal-transverse wave.  The phase
velocity of $lt_1$-mode is always subluminal while $lt_2$-mode
  is superluminal at frequencies smaller than the characteristic
  frequency $\omega_1$ (where the mode touches the $kc$ line, see
  equation~\ref{characteristic} defined later) and subluminal at
  frequencies above $\omega_1$. The dispersion relation of the $t$
mode at radio frequencies has one subluminal branch which is always
transverse nature. This mode is called the $t$-mode.  Let us underline
that polarisation vectors of linearly polarized X-and O-modes are
defined by the plane of $\vec k$ and the local magnetic field
$\vec{B}$, while the polarisation vector of observed radio waves are
parallel or perpendicular to the plane of curved dipolar field
lines. Therefore, any preferred emission mechanism should address the
problem of how the linear polarization of the plasma X and O-mode
emerges from the plasma as electromagnetic waves with the observed
polarization direction. In the rest of the paper when O-mode is
mentioned it stands for combined $lt_1$ and $lt_2$ mode, and the $X$
mode stands for the $t$ mode.

Several theoretical studies
(\citealt{1998MNRAS.301...59A,2000ApJ...544.1081M,
  2004ApJ...600..872G,2018MNRAS.480.4526L,2020MNRAS.497.3953R})
demonstrate that the observational constraints of pulsar radio
emission location and linear polarization can be explained by coherent
curvature radiation (CCR) due to motion of charge bunches
  along curved magnetic field lines. Curvature radiation has the
  feature that the radiation is predominantly polarized in the
  magnetic field plane of the particle trajectory and a fraction
  is polarized perpendicular to the plane
  \citep{jackson_classical_1999}.  In strongly magnetized plasma CCR
  due to charge bunch can excite the $t$ and $lt_1$ mode, and since
  curvature radiation mechanism can distinguish the magnetic field
  line planes the excited $t$ and $lt_1$ mode has its polarization
  perpendicular and parallel to the magnetic field line planes
  respectively. The $t$ mode has vacuum like character and can
escape freely with its polarization being perpendicular to the
magnetic field line planes.  The $lt_1$ mode has polarization in the
plane of the magnetic field and cannot easily escape from the
plasma. Thus the CCR theory still has difficulty in explaining the
emergence of polarization mode that lies parallel to the magnetic
field line planes (see \citealt{2014ApJ...794..105M}; hereafter
MMG14).

Recently PTS20 proposed that the $lt_2$-mode can be a candidate for
pulsar radio emission.  They demonstrate via particle-in-cell
simulations that due to the time dependent and non-uniform pair
creation process across the magnetic field lines in the vacuum gap,
the induced electric field in the vacuum gap is screened accompanied
by emission of waves which are electromagnetic in nature. The
  electromagnetic waves have superluminal phase velocity and PTS20
  suggests that these waves be identified with the $lt_2$ mode of the
  plasma, and as the density in the plasma decreases with height, the
waves can decouple from the plasma as electromagnetic
radiation. However, these suggestions needs to be demonstrated and the
purpose of this is to investigate if the $lt_2$ mode can be a
candidate for the observed pulsar radio emission.

The organization of the paper is as follows. Section \ref{obs} gives a
summary of the observational constraints of pulsar radio emission. In
Section \ref{Properties} the properties of O-mode are reviewed. In
Section \ref{Validity} the condition required for the validity of the
dispersion relation are established. In Section \ref{ori} the linear
polarization features of the escaping wave ($t$-mode) from pulsar
plasma are established for an arbitrary fictitious mechanism and
contrasted with CCR. Our findings are summarized in Section
\ref{conclusion}.

\section{Observational constraints of pulsar radio emission}\label{obs}
\begin{table}
    \caption{Typical multiplicity ($\kappa$) of pair plasma, radius of curvature ($\rho_\mathrm{c}$), the
	ratio $b = B_\mathrm{s}/B_\mathrm{d}$ for
	different magnetic topology at the neutron star surface.}
    \centering
    \begin{tabular}{l|c|c}\hline 
        Quantity & Dipolar & Non-dipolar  \\ \hline 
            $\kappa$  & $10-10^2$   & $10^4-10^5$    \\ \\ 
            $\rho_\mathrm{c}$ (in cm)   & $10^8$  & $10^5$  \\  \\ 
            $b $     & $1$  & $10-10^2$  \\ 
                          &        &          \\ \hline 
    \end{tabular}
    \label{Typical parameters}
\end{table}
In this section we briefly summarise the observational and theoretical inputs that constrains the parameter space of pulsar plasma.

        \begin{itemize}
		\item \textbf{Location of Radio Emission region, and
                  the direction of emergent radio waves:} Most normal
                  pulsars (defined as pulsars with periods $P$ longer
                  than about 50 milliseconds) are highly linearly
                  polarized and the linear polarization position angle
                  (PPA) across the pulse follows a characteristic
                  S-shaped curve. This PPA behaviour can be explained
                  by the rotating vector model (RVM,
                  \citealt{1969ApL.....3..225R}) according to which
                  the linear polarization vector traces the change in
                  the dipolar magnetic field line planes as the pulsar
                  radio emission beam sweeps past the line of sight of
                  the observer. As per the RVM the steepest gradiant
                  (SG) or inflexion point of the PPA traverse is
                  associated with the fiducial magnetic field
                  plane. In several pulsars two parallel PPA tracks
                  following the RVM are seen. These tracks are
                  separated by about 90$^{\circ}$, and is commonly
                  known as orthogonal polarization modes (OPM).

If the radio emission arises close to the neutron star surface then
due to the effects of abberation and retardation the center of the
total intensity pulse profile leads the SG point by a certain amount
depending on the radio emission height. This effect is observed in
several normal pulsars and the radio emission heights is constrained
to be a few hundred km above the neutron star surface, or below 10\%
of the light cylinder radius
(\citealt{1997A&A...324..981V,1991ApJ...370..643B,2004A&A...421..215M,
  2008MNRAS.391.1210W,2017JApA...38...52M,2020MNRAS.491...80P}).

Further, observations can constrain the direction of the emerging
polarization with respect to the dipolar magnetic field planes of the
pulsar.  In the case of the Vela pulsar using a combination of X-ray
image of the pulsar wind nebula and radio polarization
\cite{2001ApJ...549.1111L} found that the emerging radiation is
perpendicular to the magnetic field line planes. For several other
pulsars, by measuring the absolute PPA and the direction of the
absolute proper motion PM, the quantity $\mid$PM-PPA$\mid$ shows a
distribution around $0^{\circ}$ and $90^{\circ}$ ( see
e.g. \citealt{2005MNRAS.364.1397J,2013MNRAS.430.2281N,2012MNRAS.423.2736N,2015ApJ...804..112R,2015MNRAS.453.4485F}).
These observations are interpreted as evidence for the emerging
electric field being either parallel or perpendicular to the magnetic
field line planes that correspond to the OPM's.

\item \textbf{Highly polarized single pulses:} Average pulse profiles
  are formed by averaging hundred to thousands of single pulses,
  during which it can average out the short timescale features that
  govern the pulsar emission mechanism. Characterization of single
  pulses are hence more useful to obtain constraints for pulsar
  emission mechanism.  Single pulses from normal pulsars are composed
  of subpulses and often the subpulses are observed to be highly
  polarized, sometimes being linearly polarized close to
  100\%. \cite{2009ApJ...696L.141M} showed that when such highly
  polarized subpulses are seen, their underlying PPA follow the RVM of
  the average profile. Further MMG14 showed that
  these kind of highly polarized subpulses are associated with both
  the OPM in pulsars.  These highly polarized pulses are free from
  magnetospheric depolarization effects, and since they follow the
  RVM, their polarization direction are parallel or perpendicular to
  the magnetic field planes. Thus, any radio emission mechanism should
  be able to predict the features of highly polarized single pulses.
  
\item \textbf{Constraints on surface magnetic field and particle flow}
  The pulsar radio emission is broadband in nature and arises from
  regions of open dipolar ambient magnetic field lines located a few
  hundred km above the neutron star surface
  (\citealt{2004A&A...421..215M}). However, several lines of evidence
  suggest the presence of strong non-dipolar surface magnetic fields
  at the polar cap region.  For example radio emission from the long
  period pulsar J2144-3933 (\citealt{1999Natur.400..848Y}), can only
  be explained by efficient pair creation process which can be
  achieved by strong multipolar surface magnetic field (see
  \citealt{2001ApJ...550..383G,2020MNRAS.492.2468M}).  Also the single
  pulse phenomenon of subpulse drifting show significant variety which
  can only be explained by invoking surface multipolar magnetic fields
  ( \citealt{1975ApJ...196...51R,2020MNRAS.496..465B}).  Combined
  radio and X-ray observations can also constrain the surface magnetic
  fields.  The sparking discharges due to pair creation at the polar
  cap gives rise to a backflow of accelerated charges which bombard
  the polar cap.  This heats the polar cap (with radius
  $r_\mathrm{p}$) up to a temperature $T \sim$ few million kelvin and
  it starts emitting thermal X-rays. The X-ray luminosity provides an
  estimate of the size of the polar cap ($L_\mathrm{X} = \sigma T^{4}
  \times \pi r^2_\mathrm{p}$). X-ray studies for normal period radio
  pulsars have consistently found the polar cap sizes to be smaller
  than expected from a purely dipolar magnetic field at the polar
  cap. Using flux conservation argument, the strength of the magnetic
  field at the neutron star surface has been found to be around ten to
  hundred times stronger than the the global dipole field (Table 1 of
  \citealt{2017JApA...38...46G}). Besides, recent studies like that of
  \citealt{2019ApJ...887L..21R} have found evidence of non-dipolar
  surface magnetic field on millisecond pulsars.  In several pulsars
  there is a phase off-set between the pulsed thermal X-ray hotspot
  emission from polar cap and the radio pulse.  This phase off-set is
  larger than that is expected from a magnetic field topology which
  retains its dipolar character right up to the surface. The presence
  of strong multipolar surface magnetic fields need to be invoked to
  explain this phase offset (see
  \citealt{2017ApJ...835..178S,2019MNRAS.489.4589A,2020MNRAS.491...80P}).

	    Another consequence of the presence of non-dipolar surface
            magnetic field is that it enables efficient pair creation,
            and as a result a high density plasma pulsar wind is
            produced. Studies of pulsar wind nebula (see
            \citealt{2007ApJ...658.1177D,2011ASSP...21..624B}) has
            been used to constrain the density of the plasma, and it
            is found that the density is about 10$^4-10^5$ times more
            than the corotational Goldreich-Julian density
            $n_\mathrm{GJ}$ ( \citealt{1969ApJ...157..869G}).
        \end{itemize}
The evidence above suggests although the surface magnetic field has a
highly non-dipolar character, the higher order magnetic moments of
this configuration rapidly decays with distance such that only the
dipolar topology exists at the radio emission zone. This behaviour can
be understood in terms of the model by \citealt{2002A&A...388..235G}
(hereafter GMM02). In this model, the surface magnetic field is
modelled as a superposition of a global dipole field ($B_\mathrm{d}$)
and a crust-anchored local field ($B_\mathrm{s} = b B_\mathrm{d}$)
where $b \sim 10-100$ such that the magnetic moment of crust-anchored
field $\mu_{m}$ is much lower than the magnetic moment of the global
field $\mu_\mathrm{d}$. The contribution due to the crust-anchored
field rapidly decays with distance from the neutron star such that by
around 10 km above the surface the ambient magnetic field topology is
completely dipolar (see fig. 2 of GMM02). In this model, the field
line curvature $\rho_\mathrm{c}$ at the surface is dictated by the
local field whereas the the field line curvature $\rho_\mathrm{c}$ at
the radio emission region is dictated by the global field.

Copious pair plasma in pulsars are produced near the pulsar polar cap
where an inner accelerating region with strong unscreened electric
field exists (\citealt{1975ApJ...196...51R}). In this region high
energy photons undergo magnetic pair creation, and a high energy
primary beam relativistically moving outwards with lorentz factors of
$\sim 10^6$ can be produced. This beam can further radiate in curved
magnetic field resulting in pair cascade, which leads to the
production of the dense pair plasma having lorentz factors of about
$\sim 10^2$.  The number density of the pair plasma $n_\mathrm{s}$
exceeds the co-rotational density by a multiplicity factor $\kappa$
(\citealt{1971ApJ...164..529S}), i.e.  $n_\mathrm{s} = \kappa
n_\mathrm{GJ}$.  Let us note that pair cascade simulations
(\citealt{2002ApJ...581..451A,
  2001ApJ...560..871H,2007MNRAS.382.1833M,2015MNRAS.447.2295S,2019ApJ...871...12T})
show that smaller value of $\rho_\mathrm{c}$ on the surface increases
the pair multiplicity and vice versa. A representative value for the
multiplicity and the field line curvature are shown in Table
\ref{Typical parameters}.

In terms of pulsar parameters for a neutron star of radius
  $R_\mathrm{NS}$ the co-rotational number density at a radio emission
  height $r$ is $n_\mathrm{GJ} \sim 7 \times 10^{10}
  (\dot{P_{-15}}/P)^{0.5} (R_\mathrm{NS}/r)^3 $ cm$^{-3}$ where $P$ is
  the pulsar period in seconds and $\Dot{P}_{-15}$ is the spin-down
  rate normalized to $10^{-15}$ seconds per second.  The plasma
  frequency is $\omega_\mathrm{p} = \sqrt{ 4 \pi n_\mathrm{s}
    e^2/m_e}$ and can be written as,

\begin{equation}
    \omega_\mathrm{p} = 10^{10} \left( \frac{\kappa}{10^{5}} \right)^{1/2}
\left(\frac{\Dot{P}_{-15}}{P}\right)^{1/4} \left( \frac{500 \; \text{km}} {r} \right)^{3/2} 
	\text{rad s$^{-1}$} 
\label{plasma_freq}
\end{equation}

For typical pulsar parameters $r = 500$ km, $\kappa = 10^5$,
  $P=1$ s, $\Dot{P}_{-15} = 1$ , and $R_\mathrm{NS} = 10$ km the
  values are $n_\mathrm{GJ} \sim 5.5 \times 10^{5}$ cm$^{-3}$,
  $n_\mathrm{s} \sim 5.5 \times 10^{10}$ cm$^{-3} $ and
  $\omega_\mathrm{p} \sim 10^{10}$ rad s$^{-1}$.

\section{Properties of \lowercase{lt}$_{2}$ mode in strong ambient magnetic field}\label{Properties}

\begin{figure}
	\includegraphics[width=\columnwidth]{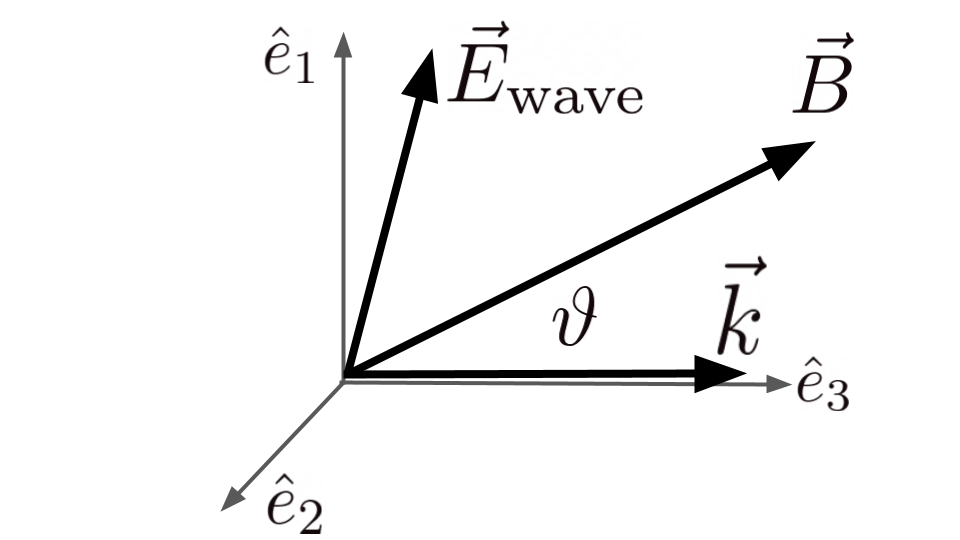}
    \caption{ Cartesian co-ordinate system
      ($\hat{e}_1,\hat{e}_2,\hat{e}_3$) adopted from \citealt{1991MNRAS.253..377K}. The wave
      vector $\vec{k}$ is taken along $\hat{e}_3$. The angle between
      the wave vector $\vec{k}$ and the dipolar ambient pulsar
      magnetic field $\vec{B}$ is $\vartheta$. The
      $k-B$ plane is confined to $\hat{e}_1 - \hat{e}_3$ plane. The
      electric field of the $lt_{2}$ waves is confined to the $k-B$
      plane.}
    \label{Ortho_coord}
\end{figure}

\begin{table}
    \centering
	\caption{Limiting values at the cut-off frequency
          $\omega_\mathrm{o}$ and the characteristic plasma frequency
          $\omega_{1}$ for $lt_2$ mode for parallel refractive index
          $n_{\parallel}$, the dimensionless variable $\chi^2$ and the
          group velocity $\Vec{\mathrm{v}}_\mathrm{g}$. Here,
          $\vartheta$ is the angle between $\vec{k}$ and $\vec{B}$ ,
          and $\Theta $ is the angle between the wave electric
          field $\vec{E}_\mathrm{wave}$ and the wave vector $\vec{k}$. The unit vector $\hat{b}_{\parallel}$ and $\hat{b}_{\perp}$ are directed along and perpendicular to the ambient magnetic field $\Vec{B}$ respectively.
        }
    \begin{tabular}{l|l|l} \hline 
        Quantity     &      $\omega \rightarrow \omega_\mathrm{o}$  & $\omega \rightarrow \omega_\mathrm{1}$  \\ \hline 
         $n_{\parallel}$ &        $0$ & $ \sqrt{1 - n^2_{\perp}}$  \\ \\ 
        $\chi^2$     & $1 - n^2_{\perp}$  &  $0$      \\  \\ 
        $\Vec{\mathrm{v}}_\mathrm{g}/c$    & $ (1 - n^2_{\perp}) \mathrm{\beta}_\mathrm{o} \; \hat{b}_{\parallel} + n_{\perp} \; \hat{b}_{\perp} $  &   $ n_{\parallel} \; \hat{b}_{\parallel} + n_{\perp} \; \hat{b}_{\perp} $ \\ \\
         $\Theta $  &   $\vartheta$ & $ {\pi}/{2} - \vartheta$  \\ 
        \hline 
    \end{tabular}\label{Limiting values}
\end{table}

The region close to the neutron star from where the pulsar radio
emission originates, the ratio of the plasma frequency to the
cyclotron frequency in terms of pulsar parameters is,
${\omega_\mathrm{p}}/{\omega_\mathrm{B}}=2\times
10^{-4}\times\kappa^{0.5}\Re^{1.5}\left( {P^{3}}/{\dot
  {P}_{15}}\right) ^{0.25} \ll 1$, where $\Re = r /R_\mathrm{LC}$ is
the ratio of emission height $r$ to the light-cylinder radius
$R_\mathrm{LC}$.  Under such condition the approximation of pair
plasma in infinite magnetic field is valid and the plasma flow can be
assumed to be strictly one-dimensional. The corresponding dispersion
relation of the O-mode of the homogeneous pair plasma in the observer
frame of reference (hereafter OFR), is given by (see Eq. 48 of
\citealt{1986ApJ...302..120A})

\begin{equation}
    \left( \omega^2  - k^2_{\parallel}c^2\right) \left[ 1 - \frac{1}{2} \sum_{\alpha}  \; \omega^2_\mathrm{p\alpha} \int^{+\infty}_{-\infty} \; dp \; \frac{f^{(0)}_{\alpha}}{\gamma^3 (\omega -  k_{\parallel} \mathrm{v} )^2}\right] - c^2k^2_{\perp} = 0  \label{O_disp}
\end{equation}
where the index $\alpha$ denotes the electron and the positron
distribution functions in the plasma. Here $k_{\parallel}$ and
$k_{\perp}$ denote components of the wave vector $\Vec{k}$ parallel
and perpendicular to the ambient pulsar magnetic field $\Vec{B}$. For
simplicity we can use the cold-plasma approximation, where the
particle distribution functions can be replaced by
delta-functions. Additionally, we assume the electron and the positron
distribution functions to be co-incident. All plasma particles are
assumed to move with the Lorentz factor $\gamma_\mathrm{s}$. Under
these simplifying assumptions, the dispersion relation of the O-mode as given by equation ($\ref{O_disp}$) reduces to the form
\begin{equation}
     \left( \omega^2 - k^2_{\parallel}c^2 \right) 
	\left[ 1 -  \frac{\omega^2_\mathrm{p}}{\gamma^3_\mathrm{s}(\omega - k_{\parallel} \mathrm{v}_\mathrm{o} )^2}\right] - k^2_{\perp}c^2 = 0 
\label{DR_orthogonal}
\end{equation}
where $\mathrm{v}_\mathrm{o}/c =\beta_\mathrm{o} = (1 - 1/\gamma^2_\mathrm{s})^{0.5}$.
The above equation has two branches, and the $lt_1$ mode correspond to
the case $\omega/k < c$ and $lt_2$ mode for which $\omega/k > c$.

\begin{figure*}
\begin{tabular}{cc}
\includegraphics[scale=0.5]{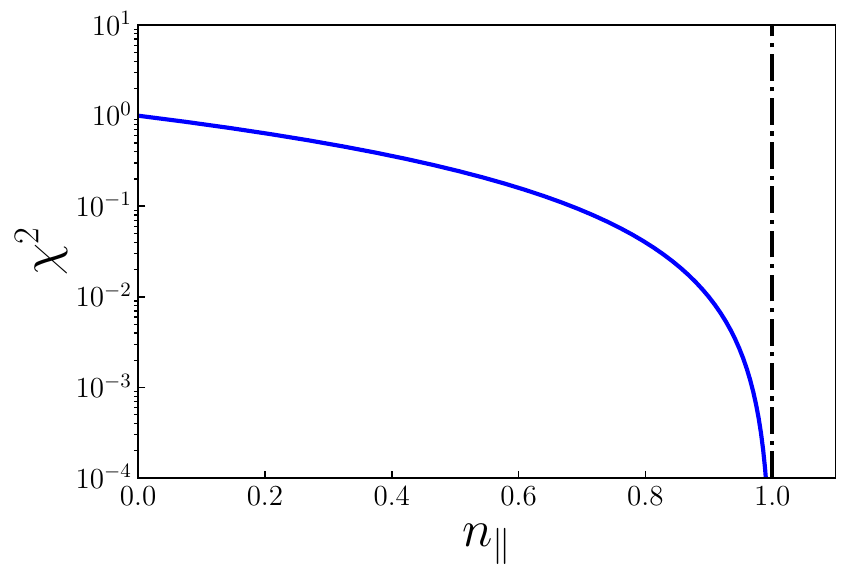} &
\includegraphics[scale=0.5]{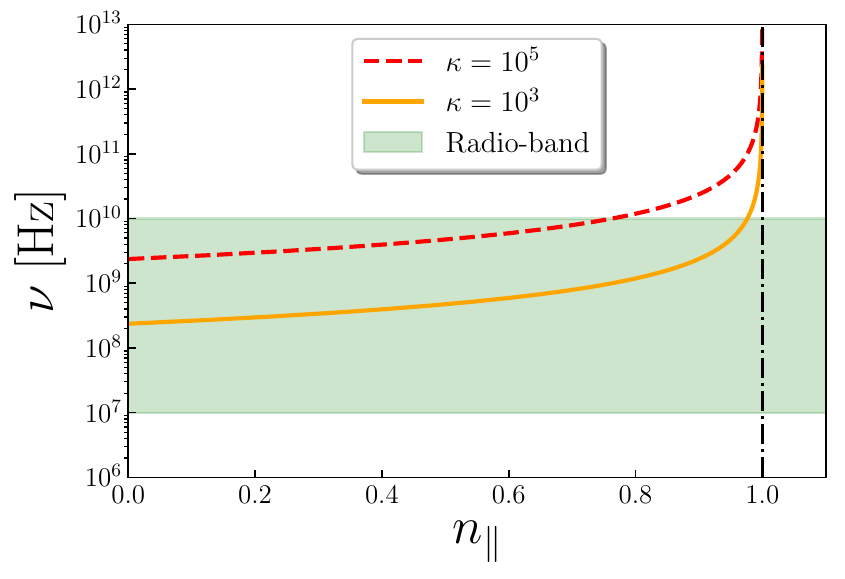} \\ (A)
Dimensionless dispersion relation for $n_\perp = 0.01$ & (B)
Dispersion relation above polar cap
\\ \includegraphics[scale=0.5]{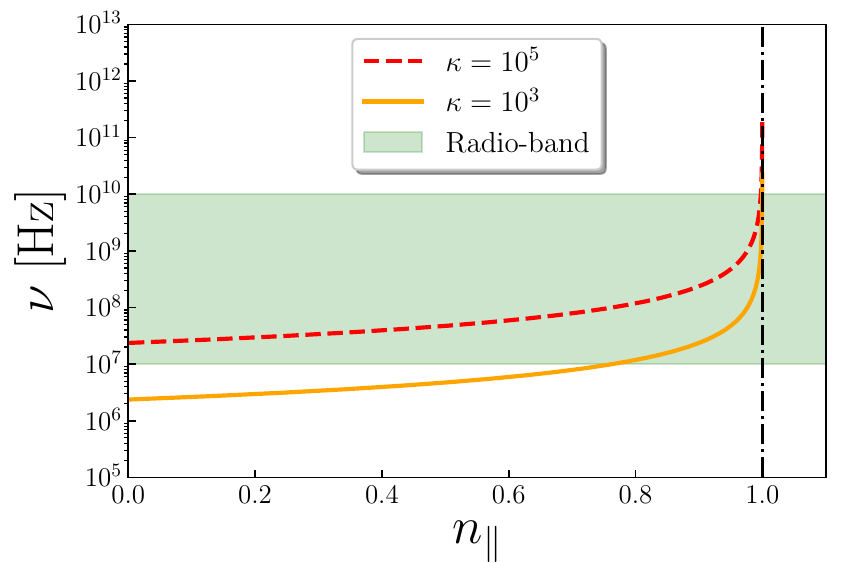} &
\includegraphics[scale=0.5]{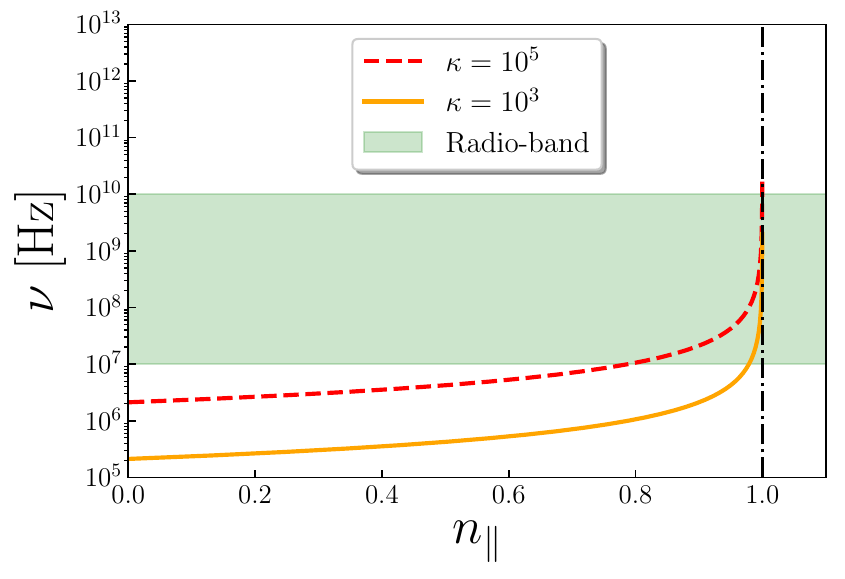} \\ (C) Dispersion
relation at $r/R_\mathrm{NS}$ = 10 & (D) Dispersion relation at
$r/R_\mathrm{NS}$ = 50 \\
\end{tabular}
\caption{The figure shows the superluminal dispersion relation for
  bulk Lorentz factor $\gamma_\mathrm{s} =100$
  at small angle of excitation ($n_\perp = 0.01$) and the radio-band
  (shown in light green) at various distances from the neutron star
  surface. The dot dashed black line in all panels shows the
  asymptotic refractive index of unity for superluminal
    $lt_{2}$ mode. The graphs are shown for the last open field line
  of an aligned rotator of period $P = 1$ sec and multiplicity $\kappa
  = 10^{5}$ (shown in dashed red line) and $10^{3}$ (shown in
    solid yellow line). \textbf{Top}: Panel (A) shows the
  dimensionless dispersion relation for small angle of propagation
  ($n_\perp = 0.01$). Panel (B) shows that a major fraction of the
  radio-band occupies the stopband just above the polar cap for a
  highly non-dipolar surface magnetic field
  ($b=10$). \textbf{Bottom:} Panel (C) shows that for $\kappa =
    10^{5}$ even 100 km above the surface the frequencies below 50
  MHz still occupies the stopband while frequencies up to 100 MHz
  lies close to the cut-off. Panel (D) shows that for $\kappa
    = 10^{5}$ only around 500 km above the surface the $lt_{2}$
  dispersion relation describes the full radio-band. Both
    Panels (C) and (D) show that for $\kappa = 10^{3}$ the radio-band
    lies above the stopband even at distances of around 100 km above
    the neutron star surface. }
\label{super_radioband}
\end{figure*}
To describe the waves we adopt co-ordinate system by \citet{1991MNRAS.253..377K} as shown in
Fig.\ref{Ortho_coord}, where the unit vectors $\hat{e}_1$,$\hat{e}_2$
and $\hat{e}_3$ are orthogonal to each other. The wave vector
$\vec{k}$ is along $\hat{e}_3$ and the ambient pulsar magnetic field
$\vec{B}$ lies in the $\hat{e}_{1} - \hat{e}_{3}$ plane.  The electric
field $\vec{E}_\mathrm{wave}$ of the O-mode lies in the $k-B$
plane. The components of $\vec{E}_\mathrm{wave}$ parallel and
perpendicular to the wave vector ($\vec{k}$) is $E_{3}$ and $E_{1}$
respectively. If $\Theta$ is the angle that the wave electric
  field of O-mode makes with the wave vector $\vec{k}$ we have,
\begin{equation}
    \tan \Theta =
\frac{E_{1}}{E_{3}} \label{tantheta}
\end{equation}

The electric field ($E_{2}$) of the t-mode is along $\hat{e}_{2}$. For
an angle $\vartheta$ between the wave vector $\Vec{k}$ and the ambient
pulsar magnetic field $\vec{B}$, the components of $\Vec{k}$ along and
perpendicular to $\Vec{B}$ are represented as $k_{\parallel} = k \cos
\vartheta$ and $k_{\perp}=k\sin\vartheta$, such that the parallel and
perpendicular refractive index can be defined as $n_{\parallel} =
k_{\parallel} c/\omega$ and $n_{\perp} = k_{\perp} c/\omega$
respectively \footnote{ Note that the difference between the
    subscripts $1-3$ that refers to components with respect to wave
  vector $\vec{k}$ and the $\parallel, \perp$ that refers to
  components with respect to the ambient pulsar magnetic field
  $\vec{B}$. }.  A plasma wave of frequency $\omega$ can escape from
the plasma only if the waves achieve vacuum-like propagation ($\omega
= kc$) so as not to suffer refraction, acquire a transverse character
($E_{3}=0$) so as to acquire electromagnetic character, and is
perpendicular to the ambient magnetic field ($E_{\parallel} = 0$) so
as not to set up any plasma current along $\vec{B}$ so as to decouple
from the plasma particles. The $t$-mode has $\omega = kc$ for all
$\omega\ll \omega_\mathrm{B}$. Besides, since $E_{2}$ is perpendicular
to the $k-B$ plane, $t$-mode automatically satisfies $E_{3} = 0$ and
$E_{\parallel} = 0$. Thus, $t$-mode can freely escape from the plasma
at the broadband radio frequencies.

For strictly parallel propagation ($\vartheta = 0$), the $lt_2$ mode admits the purely
longitudinal Langmuir ($L$) mode. The $\vec{E}_\mathrm{wave}$ of the $L$ mode is directed along the $\vec{k}$ viz., $E_{1} = 0$. 
Two characteristic frequencies can be obtained from the dispersion
relation of the $L$ mode. First, the L-mode has a cut-off frequency
$\omega_\mathrm{o}$ at the long wavelength limit ($k \rightarrow 0$)
given by
\begin{equation}
    \omega_\mathrm{o} = \frac{\omega_\mathrm{p}}{\gamma_\mathrm{s}^{3/2}} \label{cut-off}
\end{equation}
Secondly, a characteristic plasma frequency can be identified where
the L-mode touches the $\omega = kc$ line given by
\begin{equation}
    \omega_{1} = 2 \omega_\mathrm{p} \sqrt{\gamma_\mathrm{s}} \label{characteristic}
\end{equation}

For strictly perpendicular propagation ($\vartheta = \pi/2$), the
$lt_2$ admits the purely transverse and electromagnetic $TE$ mode.
The $TE$ mode has the same cut-off frequency $\omega_\mathrm{o}$ as
the L-mode and the mode remains superluminal for all frequencies
$\omega > \omega_\mathrm{o}$. The $\vec{E}_\mathrm{wave}$ of the $TE$
mode is aligned along the ambient magnetic field $\vec{B}$ with
$E_{\parallel} = E_{3} \neq 0$. Hence, even the purely transverse $TE$
mode cannot escape from the plasma.

For any intermediate angle $0<\vartheta< \pi/2$, the $lt_2$ mode has a
cut-off frequency at $\omega_\mathrm{o}$ and the longitudinal
component of the electric field along the wave vector ($E_{3}$)
dominates for frequencies near cut-off. The situation changes at
higher frequencies, where the perpendicular component $E_{1}$
dominates. Thus, an important feature of the $lt_2$-mode is that ratio
of longitudinal and transverse $E_1/E_3$ which depends on the angle
$\vartheta$ between the local magnetic field $\vec{B}$ and
$\vec{k}$. With increase of $\vartheta$ the transverse components
increase, but at the same time the phase velocity also increases and
the branch stays in superluminal area.  This could be changed if
${\omega_\mathrm{p}}/{\omega_\mathrm{B}}$ is near to unity, but this
condition never can be reached in the pulsar magnetosphere (see Eq. 1
of MMG14).

In pulsar plasma, owing to the ultra-relativistic nature of the plasma
particles excitation angles are generally considered to be small, as
they are comparable to $1/\gamma_{s}$ beam of the plasma particles,
thus $\vartheta \approx {1}/{\gamma_{s}}\sim 0.01$ rad, (using
$\gamma_{s} =100$).

\subsection{The dispersion curves and escape of $lt_2$ mode at small $\vartheta$} \label{disp curves} 

Note that equation (\ref{DR_orthogonal}) can be arranged to give the form

\begin{eqnarray}
	\chi^2 & = \frac{\omega^2_\mathrm{p}}{\gamma^3_\mathrm{s} \omega^2} &  = 
	\left( \frac{1 - n_{\perp}^2 - n^2_{\parallel}}{1 -n^2_{\parallel}}\right) (1 - n_{\parallel} \mathrm{\beta}_\mathrm{o})^2
	\label{chi}
\end{eqnarray}

where $\beta_\mathrm{o} = \mathrm{v}_\mathrm{o} /c$.  For the $lt_2$
mode $n_{\parallel}$ and $n_{\perp}$ are less than unity and further
for small angle of excitation $n_{\parallel}\gg n_{\perp}$.  For a
given $n_{\perp}$,the parallel refractive index $n_{\parallel}$
becomes the variable. It can be varied in the range $0 < n_{\parallel}
< 1$.  This gives $\chi^2$ as a function of $n_{\parallel}$ and is
referred the dimensionless dispersion relation of the $lt_2$ mode.
The $\chi^2$ so obtained can be converted to the angular frequency
($\omega$) using $ \omega = \sqrt{{\omega_\mathrm{p}^{2}}/{\chi^2
    \gamma^3_\mathrm{s}}}$. This transformation gives $\omega$ as a
function of $n_{\parallel}$ to obtain the dispersion relation. It must
be noted that for a given $\chi^2$ and Lorentz factor
$\gamma_\mathrm{s}$, the angular frequency $\omega$ (where frequency
$\nu = \omega/ 2 \pi$) scales only as $\omega_\mathrm{p}$.

 Fig. ~\ref{super_radioband} shows the $lt_2$ dispersion relation assuming a stationary plasma flow. Panel (A) shows the dimensionless
  dispersion relation of the $lt_2$ mode for small angle
  excitation. Panels (B), (C) and (D) of Fig.~\ref{super_radioband}
  shows the dimensional $lt_{2}$ dispersion relation for $\kappa =
  10^{5}$ (shown in dashed red) and $\kappa = 10^{3}$ (shown in solid
  yellow) for small $\vartheta$.  The $n_{\parallel} = 0$ in all the
  panels corresponds to the the cut-off frequency as defined in
  equation (\ref{cut-off}). For a constant $\gamma_\mathrm{s} = 100$,
  the dispersion relation is solely a function of $\omega_\mathrm{p}
  \propto \sqrt{\kappa}$.  The $lt_{2}$ mode cannot admit travelling
  wave solutions below the cut-off frequency which represents the
  stopband. As seen from Panel (B) just above the polar cap (with a
  strong non-dipolar magnetic field component) the $lt_{2}$ dispersion
  relation is valid for frequencies $\nu \gtrsim 2$ GHz for $\kappa =
  10^{5}$ and $\nu \gtrsim $ 200 MHz for $\kappa = 10^{3}$.
  Panel (C)
  shows that by around 100 km above the surface the full radio-band
  occupies the $lt_{2}$ dispersion relation for $\kappa = 10^{3}$
  while frequencies below $50$ MHz still occupies the stopband for
  $\kappa = 10^{5}$. 
   Panel (D) shows that by 500 km from the surface
  the full radio-band overlaps with the $lt_{2}$ mode for both $\kappa
  = 10^{3}$ and $\kappa = 10^{5}$. Thus, just above the polar cap the
  range of frequencies admitted depends strongly on $\kappa$ and by
  500 km from the surface the full-radio band is admitted irrespective
  of $\kappa$. It must be noted that an increase in
  $\gamma_\mathrm{s}$ (not shown) also decreases the cut-off
  frequency. Thus, the full radio-band above polar cap can be
  accommodated by decreasing $\kappa$ and increasing
  $\gamma_\mathrm{s}$ provided the plasma flow is stationary above the
  polar cap. 

\paragraph*{}

The ratio $E_{1}/{E_3}$ can be expressed as (see
Eq. 22 of MMG14),
\begin{equation}
    \frac{E_1}{E_3} = \frac{I \; \sin \vartheta \cos \vartheta}{I \; \sin^2 \vartheta - (1 - n^2)}  
	\label{exprs1}
\end{equation}
where 
\begin{equation}
     I =  {\chi^2}/{(1 - n_{\parallel} \beta_\mathrm{o})^2} \label{I}
\end{equation}

Substituting ${I} = {\chi^2}/{(1 - n_{\parallel} \beta_\mathrm{o})^2}$ from equation (\ref{I}) and $n^2 = n^2_{\parallel} + n^2_{\perp}$ in equation (\ref{exprs1})  gives, 
\begin{equation}
\frac{E_1}{E_3} =  \frac{\chi^2 \; \sin \vartheta \cos \vartheta} {\chi^2 \sin^2 \vartheta - (1 - n^2_{\perp} - n^2_{\parallel}) (1 - n_{\parallel} \beta_\mathrm{o})^2 } 
\end{equation}
But, from the definition of $\chi^2$ in equation (\ref{chi}) we
have$(1 - n^2_{\perp} - n^2_{\parallel}) (1 - n_{\parallel}
\beta_\mathrm{o})^2 = \chi^2 (1 - n^2_{\parallel})$. This simplifies
the expression above to
\begin{equation}
    \frac{E_{1}}{E_{3}} = \frac{\sin \vartheta \cos \vartheta}{n^2_{\parallel} - \cos^2 \vartheta} \label{exprs2}
\end{equation}

Using equation (\ref{tantheta}) and equation (\ref{exprs2}) the
  ratio $E_1/E_3$ can be rewritten in the form 
\begin{equation}
\frac{E_1}{E_3} = \tan \Theta  = \frac{\sin \vartheta \cos \vartheta}{n^2_{\parallel} - \cos^2 \vartheta}  \label{exprs3} 
\end{equation}

Let $\Theta_\mathrm{esc}$ be the angle that the wave electric
  field of the $lt_{2}$ mode makes with the wave vector $\vec{k}$ at
  the escape point. At small $\vartheta$ from Table (\ref{Limiting
    values}) we obtain the asymptotic parallel refractive index
  $n_{\parallel} \rightarrow \sqrt{1 - n^2_{\perp}} \rightarrow 1$
  which when substituted in equation (\ref{exprs3}) gives,
\begin{equation}
    \tan \Theta _\mathrm{esc} = \lim_{n_{\parallel} \rightarrow 1} \frac{\sin \vartheta \cos \vartheta}{n^2_{\parallel} - \cos^2 \vartheta} = \cot \vartheta 
\end{equation}
which gives  
\begin{equation}
    \Theta _\mathrm{esc} \rightarrow \pi/2 - \vartheta \label{Theta_esc}
\end{equation}
The equation (\ref{Theta_esc}) shows that the transverse character of
the $lt_{2}$ mode increases for smaller $\vartheta$.

It must also be noted that the components of $lt_2$ electric field
parallel and perpendicular to the ambient magnetic field can be found
using the following transformation
\begin{eqnarray}
    &\ E_{\parallel} = E_{3} \cos \vartheta + E_{1} \sin \vartheta  \label{Epara}\\
    &\ E_{\perp} = - E_{3} \sin \vartheta + E_{1} \cos \vartheta  \label{Eperp} 
\end{eqnarray}
The condition for decoupling of the wave from plasma in strong magnetic field (i.e. 
charges confined to move in one dimension along magnetic field lines) requires $E_{\parallel} = 0$.
Using $E_{\parallel} = 0$ in equation (\ref{Epara}) gives, 
\begin{equation}
    \frac{E_{1}}{E_{3}} = - \cot \vartheta,  \label{expr2} 
\end{equation}

Using equation (\ref{expr2}) in equation (\ref{exprs1}) and simplifying we get,
\begin{equation}
    n^2 = (1 - 2 I \; \sin^{2} \vartheta) 
\end{equation}
The condition of decoupling must be simultaneous with the wave
reaching vacuum-like propagation ($n = 1$). Substituting $n = 1$ and
the expression for $I$ gives
\begin{equation}
    \left[ \frac{\chi^2_\mathrm{decoup}}{(1 - n_{\parallel} \beta_\mathrm{o})^2} \right] \; \sin^2 \vartheta \rightarrow 0    \label{expr3}
\end{equation}
For small $\vartheta \approx 1/\gamma_\mathrm{s}$, we have $\sin \vartheta \approx \vartheta \approx 1/ \gamma_\mathrm{s}$ and $(1 - n_{\parallel} \beta_\mathrm{o}) \approx (1 - \beta_\mathrm{o}) \approx 1/\gamma_\mathrm{s}(1 + \beta_\mathrm{o}) \approx 1/ 2 \gamma_\mathrm{s}$. Using these simplifications and the expression of $\chi^2$ from equation (\ref{chi}), the equation (\ref{expr3}) reduces to the form,
\begin{equation}
    \chi^2_\mathrm{decoup} = \frac{\omega^2_\mathrm{p}}{\gamma^3_\mathrm{s} \omega^2_\mathrm{esc}} = \frac{\omega^2_\mathrm{o}}{\omega^2_\mathrm{esc}} \rightarrow 0 \label{expr4}
\end{equation}

Thus, decoupling
happens at very high frequencies ($
\omega_\mathrm{esc} \gg \omega_\mathrm{o}$ ).  Now, $\chi^2$ at the characteristic plasma
frequency $\omega_{1}$ as defined in equation (\ref{characteristic}) is given by
\begin{equation}
    \chi^2_{\omega_1} \approx \frac{1}{\gamma^4_\mathrm{s}} \geq \chi^2_\mathrm{decoup}  \label{expr5}
\end{equation}

Comparing equation (\ref{expr4}) and equation (\ref{expr5}) we get,
\begin{equation}
    \omega_\mathrm{esc} \geq \omega_{1} \label{expr6} 
\end{equation}
Note the equality condition can be used only for large
$\gamma_\mathrm{s}$. At distances of 500 km from the surface using equation ($\ref{plasma_freq}$) and equation ($\ref{characteristic}$) we have 
$\omega_\mathrm{1} \sim 2 \times 10^{11} $ rad s$^{-1}$ for $\gamma_\mathrm{s}
= 100$. It implies that even in principle decoupling of $lt_{2}$ mode
at around 500 km occurs at small $\vartheta$ for frequencies $\nu \gtrsim 30$ GHz for $\kappa = 10^{5}$. 

The escape of the $lt_2$ wave requires the wave to acquire a
transverse character in the plasma frame. 
It was shown by
MMG14 (see their Eq. 26) that the $lt_2$ mode
acquires almost an electromagnetic character i.e. $E_{3}\rightarrow0$
in the observer frame when the condition $\gamma_\mathrm{s} \vartheta
> 1$ is satisfied for $\omega \gtrsim \omega_{1}$. In order to
understand the physical nature of this behaviour it is convenient and instructive to
consider features of this wave in the plasma frame of reference. The plasma frame (referred to as the primed frame) moves with respect
to the observers' frame (referred to as the unprimed frame) along the
direction of the local magnetic field at the Lorentz factor $\gamma_\mathrm{s}$.  In the plasma frame the wave
electric field components along and across the magnetic field can be
described as
\begin{eqnarray}
E_{1}^{\prime} &  =E_{\parallel}^{\prime}\sin\vartheta^{\prime}+E_{\perp}^{\prime
}\cos\vartheta^{\prime}  \label{E1_prime} \\
E_{3}^{\prime} &  =E_{\parallel}^{\prime}\cos\vartheta^{\prime}-E_{\perp}^{\prime
}\sin\vartheta^{\prime} \label{E2_prime} 
\end{eqnarray}

Here
$\vartheta^{\prime}=k_{_{\perp}}^{\prime}/k_{_{\parallel}}^{\prime}$
is the ratio of the wave vector components in the plasma frame. From
the Lorentz transformation for the wave vector, it follows that
$k_{\perp}=k_{_{\perp}}^{\prime}$ and $k_{_{\parallel}}\approx \gamma_\mathrm{s}
k_{_{\parallel}}^{\prime},$ thus in the plasma frame
$\ \vartheta^{\prime} \approx \gamma_\mathrm{s} \vartheta$ and using the condition
$\gamma_\mathrm{s} \;\vartheta>1$, we get $\vartheta^{\prime}>1$. Similarly we can
estimate the components of the electric field of a wave, as
$E_{\parallel} = E_{\parallel}^{\prime}$ and $E_{\perp} \simeq \gamma_\mathrm{s}
E_{\perp }^{\prime}$, and hence the fulfilment of the condition
$E_{\perp}\gg E_{\parallel}$ (i.e. the wave is almost transverse in
the observers' frame) does not guarantee the fulfilment of the
condition $E_{\perp}^{\prime}\gg E_{\parallel }^{\prime}$ (i.e. the
wave is also almost transverse in the plasma frame).  Indeed,
estimations as in MMG14 show that $E_{\perp} \approx \gamma_\mathrm{s}
E_{\parallel},$ so it is clear from the equations  that if
$\vartheta^{\prime }$ has an intermediate value between $0$ and
$\pi/2,$ $E_{1}^{\prime}$ and $E_{3}^{\prime}$ should be the same order in the plasma frame. It must be noted that our analysis assumes a homogeneous plasma flow. The possibility of the escape of $lt_{2}$ waves under inhomogeneous plasma conditions requires further study and is outside the scope of this work. 

\subsection{Dependence of escape conditions on $\vartheta$ and $\kappa$} \label{section3.2}

We have discussed above the escape conditions for small values of $\vartheta$ which is
valid for dipolar magnetic field at the polar cap. However
in the PTS20 mechanism this angle is measured between the normal to the pair production front (i.e. the direction of the wave vector) and the ambient magnetic field, and can, in general, vary by many orders of magnitude (up to $\sim$ 1 radian), depending on the ambient magnetic field profile and on the shape of the pair production front.
In the
case of a highly non-dipolar surface field due to the small radius of 
curvature of the magnetic field lines the pair creation front can make 
large angle with respect to the
global dipolar magnetic field axis. However in this case, the wave vector 
will not necessarily be directed along the open dipolar magnetic field lines but will propagate towards the 
close magnetic field line directions. 

The presence of radio wave as traveling $lt_2$ mode in pulsar plasma also depends 
on the multiplicity $\kappa$.
The characteristic frequency $\omega_{1}$ can be lowered to occupy the radio-band if the density of the pair plasma decreases such that multiplicity $\kappa \leq 10^3$. In fact simulations 
suggests that during pair creation process only a small portion of the primary beam
attains $\kappa \sim 10^5$, while the average multiplicity of the trail has smaller 
values of $\kappa \sim 10^2 - 10^3$ (see e.g. \citealt{2010MNRAS.408.2092T}), and hence
there exist parameter space where $lt_2$ mode can have propagating solutions. However as
we will discuss in section ~\ref{Validity} an additional requirement to obtain valid 
traveling wave solutions for $lt_2$ mode is the presence of a stationary plasma is essential.

\subsection{Constraints due to Group velocity} \label{Section3.4}

The energy pumped at a given angular frequency $\omega$ is transported
at the the group velocity ($\Vec{v}_\mathrm{gr}$) of the $lt_{2}$ mode
given by (see Appendix \ref{appB})
\begin{equation*}
  \frac{1}{c} \; \Vec{v}_\mathrm{gr} (n_{\parallel}, n_{\perp}) = \mathrm{\beta}_\mathrm{gr,\parallel} \; \hat{b}_{\parallel} + \mathrm{\beta}_\mathrm{gr,\perp} \; \hat{b}_{\perp} 
\end{equation*}
where $b_{\parallel}$ and $b_{\perp}$ are the unit vectors parallel
and perpendicular to the ambient pulsar magnetic field
respectively. Here, the components $\beta_\mathrm{gr,\parallel}$ and
$\beta_\mathrm{gr,\perp}$ are given by
\begin{eqnarray}
    &\ \mathrm{\beta}_\mathrm{gr,\parallel} =  \frac{ n_{\parallel} - \frac{n^2_{\parallel} \chi^2 }{(1 - n_{\parallel} \mathrm{\beta}_\mathrm{o})^2 } + \frac{ (1 - n^2_{\parallel}) \chi^2 \mathrm{v}_\mathrm{o} } {(1 - n_{\parallel} \mathrm{\beta}_\mathrm{o})^3}   } { \Bigg\{ 1 - \frac{\chi^2}{(1 - n_{\parallel} \mathrm{\beta}_\mathrm{o})^2 } \Bigg \} + \frac{\chi^2 (1 - n^2_{\parallel}) }{ (1 - n_{\parallel} \mathrm{\beta}_\mathrm{o})^3}   }  \\
    &\ \mathrm{\beta}_\mathrm{gr,\perp} =  \frac{n_{\perp}}{ \Bigg\{ 1 - \frac{\chi^2}{(1 - n_{\parallel} \mathrm{\beta}_\mathrm{o})^2 } \Bigg \} + \frac{\chi^2 (1 - n^2_{\parallel}) }{ (1 - n_{\parallel} \mathrm{\beta}_\mathrm{o})^3}   } 
\end{eqnarray}

Thus, the plasma waves propagate at an angle $\vartheta _\mathrm{gr}$
to the local ambient magnetic field expressed as
\begin{equation}
    \tan \vartheta_\mathrm{gr} = \frac{\beta_\mathrm{gr,\perp}}{\beta_\mathrm{gr,\parallel}}  \label{grp_vel_angle}
\end{equation}

As discussed in the previous section, the pulsar magnetic field topology at the wave generation points is purely dipolar. The equation (\ref{grp_vel_angle}) shows that as a result of propagation the angle $\vartheta$ between the ambient magnetic field $\vec{B}$ and the wave vector $\vec{k}$ necessarily increases for a diverging set of magnetic field lines. Let $\vartheta_{gen}$ and $\vartheta_\mathrm{esc}$ be the angle between the wave vector $\vec{k}$ and $\vec{B}$ at the generation point $r_\mathrm{gen}$ and the escape point respectively $r_\mathrm{esc}$. Thus, if $r_\mathrm{gen}$ and $r_\mathrm{esc}$ are far away as a result of propagation we will have $\vartheta_\mathrm{esc}
\gg \vartheta_\mathrm{gr} + \vartheta _\mathrm{gen}$. As an illustrative example, consider  $r_\mathrm{gen}$ and $r_\mathrm{esc}$ to be $ \Delta r \sim$ 500 km apart. The typical radius of curvature for magnetic field lines in the inner magnetosphere at radio emission heights is $\rho_{c} \sim 10^{8}$ cm. The change in angle $\Delta \vartheta$ between wave vector and the ambient magnetic field due to propagation gives $ \Delta \vartheta \approx \Delta r/ \rho_\mathrm{c} \sim 0.5$ radians. Thus, even if $\vartheta_\mathrm{gen}$ were to be small, propagation increases $\vartheta_\mathrm{esc}$. This violates
the small angle approximation that is necessary for the plasma waves
to acquire transverse character at $r_\mathrm{esc}$. To prevent this
situation the underlying emission mechanism must ensure that the
plasma waves are generated very close to the escape point. This
requires that the general class of $lt_{2}$ models must have a local
character.

The limiting behavior of the properties of the $lt_2$ waves for small
angle of excitation at the cut-off frequency $\omega_\mathrm{o}$ and
the characteristic frequency $\omega_\mathrm{1}$ are summarized in
Table \ref{Limiting values}.

\section{Validity of dispersion relation for Externally driven model of Radio Emission}
\label{Validity}


As a particular example of an externally driven mechanism that can
excite $lt_2$ mode, we consider the recent mechanism by PTS20. The
model appeals to the direct energy transfer from the transient
electromagnetic field in the polar cap to the $lt_2$ mode due to a
time-dependent gap screening process. The mechanism works as
follows. During the pair cascade in the polar gap, the number density
$n_\mathrm{s}(t)$ of the pair plasma changes. As a result, a
time-dependent plasma frequency $\omega_\mathrm{p}(t) \propto
\sqrt{n_\mathrm{p}(t)}$ can be associated with the pair plasma. PTS20
suggests that energy associated with the transient external
electromagnetic field at the polar gap is pumped at the time-dependent
cut-off frequency $\omega_\mathrm{o}(t) \propto
\omega_\mathrm{p}(t)$. In the time interval ($0 < t < t_\mathrm{max}$)
the cut-off frequency $\omega_\mathrm{o}(t)$ spans the full radio band
(roughly $\nu \sim$ 10 MHz to 10 GHz).

While the suggestion of PTS20 is attractive, it is important to assess
if the electromagnetic radiation that is produced due to the rapidly
varying pair creation process can couple to the eigen modes of the
relativistically streaming pulsar plasma. To investigate this let us
briefly recall the non-stationary pair creation process above the
polar cap as described in the inner vaccum gap models
(\citealt{1971ApJ...164..529S}; \citealt{1975ApJ...196...51R}; hereafter RS75).  The process
starts when a high energy gamma ray photon traverses a mean free path
distance of about $h_\mathrm{g}$ before it gets converted into an
electron positron due to the presence of strong magnetic field above
the polar cap. RS75 pointed out that this
condition requires the $h_\mathrm{g}$ to be similar to the size of
polar cap radius $h_\mathrm{g} \approx r_\mathrm{pc}$.  For a purely
dipolar surface geometry, $r_\mathrm{pc,dip} \approx 100$ metres for a
pulsar with $P=1$ seconds.  For the non-dipolar surface geometry, the
polar cap radius $r_\mathrm{pc, non-dip} = r_\mathrm{pc, dip}/\sqrt{b}
\sim 30$ meters using $b = 10$. Using $h_\mathrm{g} \approx
r_\mathrm{pc,non-dip} \sim 30$ meters. RS75 suggested that the typical timescale $\tau$ associated with pair creation to be the light crossing time associated with $h_\mathrm{g}$. Thus, $\tau$ is given by
\begin{equation}
    \tau \sim h_\mathrm{g}/c = 100 \; \text{ns} = 10^{-7} \; \text{s} \label{tau}
\end{equation}
The pair creation process then undergoes a cascade until the induced electric field above the polar cap is screened completely, which happens in a time $t \approx 30- 40 \tau$ (see RS75).  The numerical simulations of \cite{2010MNRAS.408.2092T} have confirmed the same. Thus, the maximum time of discharge can be taken to be,
\begin{equation}
    t_\mathrm{max}\sim 40\tau = 4000 \; \text{nanoseconds} \label{tmax}
\end{equation}
So, in a maximum time $t_\mathrm{max} \approx 4000$ nanoseconds a dense
pair plasma can be produced above the polar cap having number density
corresponding to,
\begin{equation}\label{nmax}
  \begin{split}
    & n_\mathrm{s,max}(t = t_\mathrm{max}) = \kappa\; \frac{b B_\mathrm{d}}{P c e} \\ & \approx 10^{17} \left( \frac{1 \; \text{s} }{P}\right) \left( \frac{B_\mathrm{d}}{10^{12} \; \text{G}} \right) \left( \frac{b}{10}\right)\left( \frac{\kappa}{10^{5}} \right) \; \text{cm$^{-3}$}  
  \end{split} 
\end{equation}

The excited frequency $\omega$ at any time $t$ can be represented as
\begin{equation}
    \omega (t) = \omega_\mathrm{o} (t) = a \; \sqrt{ \left( \frac{n_\mathrm{s}(t)}{\text{cm$^{-3}$}} \right) }
\label{freq}
\end{equation}
where $a$ is a constant expressed as
\begin{equation}
	a = \sqrt{ \frac{4 \pi e^2}{m_\mathrm{e}} \; \frac{1}{\gamma^3_\mathrm{s}}}  \approx 56 \; \sqrt{\left( \frac{10^2}{\gamma_\mathrm{s}} \right)^3 }\; \text{rad s$^{-1}$} 
\label{a}
\end{equation}

It must be noted that after a pair plasma discharge is complete (i.e., the plasma settles to a stable flow), the
final cut-off frequency $\omega_\mathrm{max} = \omega_\mathrm{o} (t
= t_\mathrm{max})$ of the $lt_{2}$ mode is given by
\begin{equation}\label{cutmax}
\begin{split}
    & \omega_\mathrm{max} \approx \\ 
    &  10^{10} \sqrt{ \left( \frac{1 \; \text{s}}{\mathrm{P}} \right)\;\left( \frac{10^2}{\gamma_\mathrm{s}} \right)^3 \; \left( \frac{  \kappa} {10^{5}} \right) \; \left( \frac{B_\mathrm{d}}{10^{12} \; \text{G} } \right) \; \left(\frac{b}{10} \right) } \;  \text{rad s$^{-1}$}
\end{split}     
\end{equation}
From panel (B) of Fig. \ref{super_radioband} it can be seen that the
lower frequencies $\omega< \omega_\mathrm{max}$ occupies the
stop-band of the $lt_{2}$ dispersion relation. The $lt_{2}$ mode
cannot exist in the stop-band. The higher frequencies $\omega >
\omega_\mathrm{max}$ cannot be excited as the proposed mechanism
pumps energy only nearer the dynamic cut-off
$\omega_\mathrm{o}(t)$. As a result, the higher radio frequencies
remain inaccessible to the direct energy transfer. Thus, the external
energy transfer during gap discharge cannot excite $lt_{2}$ mode at
radio frequencies to the plasma.  Next we consider in detail as to why
the notion of plasma wave solutions at broadband radio frequencies
remains invalid for rapid pair cascade processes above polar
cap.

Using equation (\ref{freq}) the change of the excitation
frequency ($\omega$) is governed by the equation
\begin{equation}
    \frac{d \omega }{dt} = a \; \frac{d \; }{dt}  \sqrt{n_\mathrm{s}(t)} 
\label{freqev}
\end{equation}

The oscillation period ($\mathcal{T}$) associated with frequency
$\omega$ is given by
\begin{equation}
    \mathcal{T} = \frac{1}{2 \pi \omega}  
\label{timeperiod}
\end{equation}

We now track the cascade process by slicing the total discharge time
into smaller discrete time intervals $\Delta t$. The number density of
the pair plasma can be assumed to be constant within $\Delta t$. The
frequency of excitation changes by a fixed and small $\Delta \omega$
in each time interval ($\Delta t$). We assume that in the time interval $\Delta t$ the plasma frequency $\nu_\mathrm{p}$ changes by $\Delta \nu = 1$ MHz viz., a small fraction of the lowest observed radio
frequency $\nu_\mathrm{low} =  10$ MHz. This gives us the corresponding angular frequency $\Delta \omega$,   
\begin{equation}
    \Delta \omega = 2 \pi \Delta \nu \approx 10^{7} \; \text{rad s$^{-1}$}, \label{Delta_omega} 
\end{equation}
A plasma wave of frequency $\omega$ must have sufficient number of oscillations within $\Delta t$. Using equation (\ref{timeperiod}) this condition can be represented as,
\begin{equation}
	\frac{\Delta t}{\mathcal{T}} = 2 \pi \omega \Delta t \gg 1, 
\label{validity}
\end{equation}
A plasma wave can be said to be excited only if the above condition
holds. In what follows we consider two models of plasma discharge and
check for the validity of equation (\ref{validity}). The models are a linear
and an exponential plasma injection model. 

     In the linear model, the number density of the pair plasma
     increases linearly with time. This model is used in the recent
     study by PTS20. The pair plasma particles are injected at a
     constant rate $\Dot{n}$. The number density of pair particles
     changes as
    \begin{equation}
        n_\mathrm{s}(t) = \Dot{n} t  \hspace{4cm} \text{For $0< t < t_\mathrm{max}$}
\label{m1}
\end{equation}
    The constant injection rate can be estimated using $t_\mathrm{max}
    = 40 \tau$ from equation (\ref{tmax}) and using $n_\mathrm{s,max}$ from equation (\ref{nmax}) as 
    \begin{equation}\label{ndot}
        \begin{split}
        & \Dot{n} = \frac{n_\mathrm{s,max}}{t_\mathrm{max}}  \approx \\ &  2. 5 \times 10^{23}  \left( \frac{10^{-7} \; \text{s}}{\tau}\right) \left( \frac{\kappa}{10^{5}}\right)\; \left( \frac{b}{10}\right) \left( \frac{B_\mathrm{d}}{10^{12} \; \text{G}}\right) \left(\frac{1 \;  \text{s}}{P} \right) \text{ cm$^{-3}$ s$^{-1}$ } 
	   \end{split}
    \end{equation}

   Substituting equation (\ref{m1}) into equation (\ref{freqev}) gives
    \begin{equation}
        \Delta t = \frac{2 \omega \Delta \omega }{a^2 \Dot{n}} \label{tlinear}
    \end{equation}
    Substituting the above expression for $\Delta t$ into equation
    (\ref{validity}) and then using the value of $a$ from equation
    (\ref{a}) and the value of $\Dot{n}$ from equation (\ref{ndot}) we
    get,
    \begin{equation}\label{linear_condn}
    \begin{split}
        &\ \omega^2 \Delta \omega  \gg \\ &     10^{25}   \left( \frac{10^{-7} \;  \text{s}}{\tau}\right)\; \left( \frac{\kappa}{10^{5}}\right) \; \left( \frac{b}{10}\right) \; \; \left( \frac{B_\mathrm{d}}{10^{12}\; \text{G}}\right) \;\left( \frac{10^2}{\gamma_\mathrm{s}} \right)^3 \left( \frac{1 \;  \mathrm{s}}{P}\right) \text{rad$^{3}$ s$^{-3}$} 
    \end{split}     
    \end{equation}
    
   The equation above shows that the largest number of oscillations can be accommodated for $\omega = \omega_\mathrm{max}$. Using the expression for $\omega_\mathrm{max}$ from equation (\ref{cutmax}) in equation (\ref{linear_condn}),  it can be seen that $\omega^2_\mathrm{max} \Delta \omega \approx 10^{27}$ rad$^{3}$ s$^{-3}$ such that the dependence on $\kappa$ and $\gamma_{s}$ vanishes and about hundred temporal oscillations can be accommodated.

    The second model is a more realistic model, where the number
    density of the pair plasma increases exponentially
    (\citealt{1971ApJ...164..529S} ; \citealt{1975ApJ...196...51R} ;
    \citealt{2001ApJ...560..871H}; \citealt{2002ApJ...581..451A}). The
    number density of pair plasma can be modelled as
    \begin{equation}
        n_\mathrm{s}(t) = \exp{(t/\tau)} \hspace{3cm }\text{For $0 < t < t_\mathrm{max}$} 
	    \label{m2}
    \end{equation}
    Substituting equation (\ref{m2}) in equation (\ref{freqev}) gives
    \begin{equation}
        \Delta t = \frac{2 \tau \Delta \omega}{\omega}  \label{texp}
    \end{equation}
    Substituting the above expression for $\Delta t$ into equation (\ref{validity}) gives
    \begin{equation}
	     4 \pi \Delta \omega \; \tau \gg 1  \label{exp_cond1}
    \end{equation}
    From equation (\ref{tau}) we estimate  $\tau \sim 10^{-7}$ seconds, 
    and it can be seen that in
    the exponential model up to ten temporal oscillations can be completely
    fulfilled across the full broad-band radio frequencies.
    
    It must be noted that in the exponential model the condition  (\ref{validity}) for the validity of travelling wave solution as expressed in equation (\ref{exp_cond1}) does not depend on $\kappa$ or $\gamma_\mathrm{s}$. Physically, it can be understood as follows: In the exponential model plasma number density increases e-fold after each characteristic timescale $\tau$. Thus, the timescale $\Delta t$ can only be a tiny fraction of the characteristic timescale $\tau$ as shown in equation (\ref{texp}).

    It must also be mentioned during discharge the spatial scale $l =
    c \Delta t$ must be able to accommodate several ($\mathcal{N} \gg
    1$) wavelengths $\lambda$ for a plane travelling wave solution
    such that
    \begin{equation}
        \mathcal{N} = \frac{c \Delta t}{\lambda} = \frac{1}{2 \pi} \; n \omega \Delta t  \label{N}  
    \end{equation}
    Using equation (\ref{texp}) we have $ \omega \Delta t = 2 \tau \Delta \omega$. Using $\tau = 10^{-7}$ seconds from equation (\ref{tau}) and $\Delta \omega = 10^{7}$ rad s$^{-1}$ from equation (\ref{Delta_omega}) we obtain the value $2 \tau \Delta \omega = 2$ . This value when substituted in equation (\ref{N}) gives,  
    \begin{equation}
        \mathcal{N} = \frac{ n}{\pi} \left(\frac{\tau}{100 \;\text{ns}}  \right)  \approx 0.32 \;n \left(\frac{\tau}{10^{-7} \;\text{s}}  \right) \label{N2} 
    \end{equation}
     The refractive index for the $lt_{2}$ mode is $n = ck /\omega < 1$ and closer to the dynamic cut-off $n \rightarrow n_{\perp}$. 
    The $lt_{2}$ dispersion relation requires $n_{\perp} < 1$.  Thus, irrespective of $n_{\perp}$ (and hence $\vartheta$) the dynamic scale $l$ can hardly accommodate one wavelength across the entirety of the discharge process.

Thus, during pair plasma discharge plasma waves at radio frequencies
cannot be generated. It is because the time interval for which plasma
density can be assumed to be constant is too short to accommodate many
oscillations (temporal and spatial) as is necessary for travelling
plasma wave solutions to exist. Physically, the description of normal
modes in a plasma is only possible when the plasma has settled to a
stable state during which plasma frequency is constant.  The stable
state plasma supports eigen modes and admits travelling wave
solutions.  The solutions at short times are referred to as
transients. As discussed above, for very rapidly changing number
density model the very notion of oscillatory plasma wave solutions
remains inapplicable, and hence the electromagnetic waves found in the
simulation of PTS20 cannot be identified as the eigen modes of the pair
plasma. The total energy in the plasma is divided into the kinetic
energy of the plasma particles and the normal modes of plasma
oscillations. The energy pumped during the transient discharge phases
can increase only the ``temperature'' (kinetic spread of the plasma
distribution function) of the pair plasma. It is worth mentioning that 
the arguments presented in this section are not relevant for the 
externally rotation-driven pulsar emission mechanism studied by \cite{2021MNRAS.500.4549M}.

To summarize, the time-dependent pair discharge process at the polar
gap cannot pump energy into $lt_{2}$ mode at radio frequencies. Any
energy transfer during screening can only increase the momentum spread
of the particle distribution function. The full radio-band can satisfy
the $lt_{2}$ plasma dispersion relation only at few hundreds of km
away from the surface (see Fig. \ref{super_radioband}). It is at these
distances that the $lt_{2}$ mode can be excited in principle. However,
it was shown in the previous section that at these heights the
$lt_{2}$ waves cannot decouple from the plasma at radio
frequencies. Only the $t$-mode can decouple from the plasma at the
radio emission zone. In the next section we consider a fictitious
mechanism that can excite $t$ waves and analyse how the escape
conditions dictates the orientation of the escaping waves with respect
to the curved magnetic field lines.

\section{Orientation of \lowercase{$t$}-mode  at the decoupling zone}
\label{ori}

\begin{figure}[h]
    \includegraphics[width=\columnwidth]{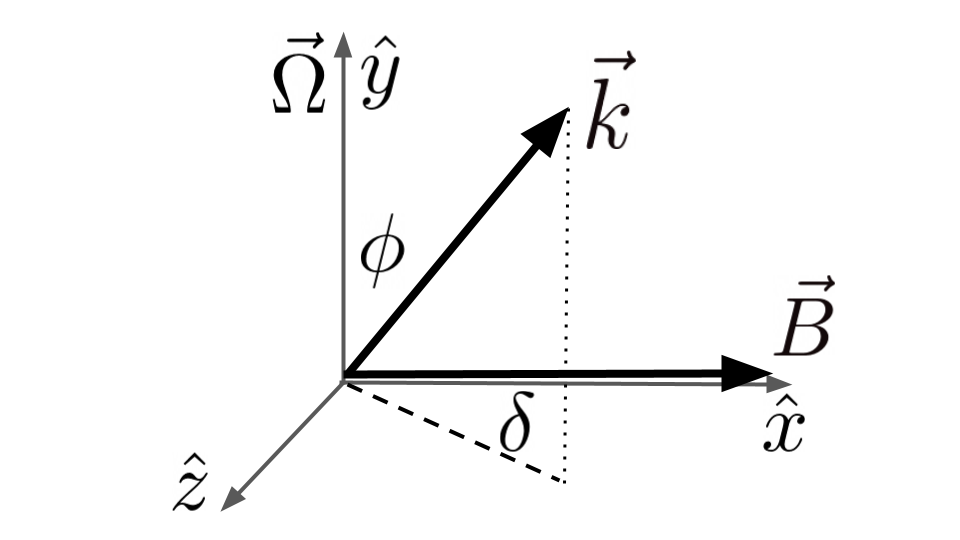}
    \caption{The figure shows the orthogonal co-ordinate system
      associated with the magnetic field plane. The $\hat{x},
      \hat{y},\hat{z}$ are the orthogonal unit vectors associated with
      the co-ordinate system. The $x-y$ plane corresponds to the
      magnetic field plane such that the rotation axis $\vec{\Omega}$
      is along the $y$-axis while the ambient magnetic field $\vec{B}$
      is along $x$-axis. The wave vector $\vec{k}$ makes an angle
      $\phi$ with the $y$-axis and the projection of $\vec{k}$ on
      $x-z$ plane makes an angle $\delta$ with the $x$ axis.}
    \label{Magnetic_field_plane}
\end{figure} 
In section \ref{Validity} we argued that the electromagnetic waves
generated due to the time-dependent pair creation process above polar
cap cannot couple to the plasma eigen modes. Further as discussed in
subsection \ref{disp curves}, we have argued using general arguments
that the $lt$ mode cannot easily escape the plasma. Only the
extra-ordinary $t$ mode can escape from the homogeneous plasma.  Here
we consider a fictitious emission mechanism which can excite $t$ mode
in pair plasma.  As mentioned in section \ref{obs} an important
observational requirement for the escaping wave to be a candidate for
pulsar radio emission is that the linear polarization of the emerging
radiation has to be parallel or perpendicular to magnetic field line
planes. A further requirement is that the emission mechanism should be
able to produce very highly linearly polarized (close to 100\%)
subpulses that follow the RVM. Hence our aim is to verify 
that if a fictitious emission mechanism 
that excites the $t$ mode can be a candidate for pulsar radio
emission. The first observational constraint requires us to find the
find the direction of the excited wave with respect to magnetic field
line planes and the second one requires to specify an averaging
process that does not lead to significant depolarization.

Let us note that at about 10\% of the light cylinder radius, where
radio emission detaches from the pulsar magnetosphere, the radius of
curvature of dipolar field lines are significantly larger than the
wavelength of radio waves. This allows us to introduce a local
cartesian co-ordinate system for the magnetic field plane as shown in
Fig. \ref{Magnetic_field_plane}. In this representation the ambient
magnetic field $\vec{B}$, the wave vector $k$ and the wave electric
field ($\vec{E}_{2}$) of the $t$-mode are given by
\begin{eqnarray}
    &\ \vec{B} = B \; \hat{x}  \label{B}\\ 
    &\ \vec{E}_{2} = E_{2x} \; \hat{x} + E_{2y} \; \hat{y} + E_{2z} \; \hat{z} \label{E2}\\
    &\ \vec{k} = k_{x} \; \hat{x} + k_{y} \; \hat{y} + k_{z} \; \hat{z}   \label{k}
\end{eqnarray}

Since for the $t$-mode $\vec{E}_{2}$ is perpendicular to the $k-B$
plane, the following two conditions must be satisfied
\begin{eqnarray}
    &\ \vec{E}_{2} \cdot \vec{B} = E_{\parallel} = 0 \label{dot1}\\
    &\ \vec{E}_{2}  \cdot \vec{k} = 0 \label{dot2}
\end{eqnarray}

Using equation (\ref{B}) and equation (\ref{E2}) in equation
(\ref{dot1}) we get,
\begin{equation*}
    E_{2x} = 0 \label{E2x}
\end{equation*}
Using the result above, the wave electric $E_{2}$ can be represented
as
\begin{equation}
    \vec{E}_{2} = E_{2y} \; \hat{y} + E_{2z} \; \hat{z}  \label{newE2} 
\end{equation}

Using equation (\ref{k}) and equation ($\ref{newE2}$) in equation
($\ref{dot2}$) we have,
\begin{equation}
    \frac{E_{2y}}{E_{2z}} = - \frac{k_{z}}{k_{y}} \label{ratio}
\end{equation}

From Fig. \ref{Magnetic_field_plane} the components of the wave vector
($\vec{k}$) can be represented as
\begin{eqnarray}
    &\ k_{y} = k \cos \phi  \label{ky}\\
    &\ k_{x} = k \sin \phi \cos \delta \label{kx} \\
    &\ k_{z} = k \sin \phi \sin \delta \label{kz}
\end{eqnarray}
Substituting equation (\ref{ky}) and equation (\ref{kz}) in equation
(\ref{ratio}) we define a dimensionless quantity $\xi$ given by,
\begin{equation}
    \xi = \frac{E_{2y}}{E_{2z}} = - \tan \phi \sin \delta \label{ratio2}
\end{equation}
where $E_{2y}$ and $E_{2z}$ are the component of the wave electric
field in the magnetic field plane and perpendicular to it. It must be
noted that the angles $\phi$ and $\delta$ cannot vary
independently. For a fixed angle of excitation $\vartheta $ between
$\vec{k}$ and $\vec{B}$, the dot product $\vec{k} \cdot \vec{B}$ must
remain constant. This requires,
\begin{eqnarray}
    \cos \vartheta  = \sin \phi \cos \delta = \text{constant} \label{kbdotprod}
\end{eqnarray}
which shows that many pairs of $(\phi, \delta)$ can satisfy the
relations above. However, for non-zero $\vartheta $ the pairs $(\phi =
0, \delta)$ and $(\phi, \delta = \pi/2 )$ are excluded. Physically,
this means that the wave vector $\vec{k}$ can trace a cone with
opening angle ($\vartheta $) centred on the ambient magnetic field
$\vec{B}$.

Now our objective is to find the condition under which we can observe
very high linear polarization of subpulses that follow the RVM in the
pulsar radio emission zone (\citealt{2009ApJ...696L.141M}). The
typical angular width of a subpulse is around $\sim 3^{\circ} \sim 6
\times 10^{-2}$ radians.  For a bulk plasma flow with
$\gamma_\mathrm{s} = 100$, the typical emission cone due to any
fictitious emission mechanism has opening angle $\vartheta \sim
1/\gamma_\mathrm{s} \sim 10^{-2}$ radians. As the observer samples a
subpulse the net emission at any point is an incoherent superposition
of the electric field due to a large number of $1/ \gamma_\mathrm{s}$
emission cones that arises from nearby field lines.  As discussed
above, for generic fictitious mechanism(s) the electric field for an
emission cone has no specific orientation with respect to the magnetic
field plane. As a result, the incoherent averaging will necessarily
lead to depolarization of the subpulse. In order to retain the high
linear polarization the emission mechanism must preferentially select
a fixed orientation of the $t$-mode electric field within $1/\gamma_\mathrm{s}$
emission cones. Further, observations suggest that the emerging radio 
waves are polarized perpendicular to the magnetic field line planes \citep[see e.g][]{2001ApJ...549.1111L}.
The only known mechanism that can distinguish between the directions
parallel and perpendicular to the magnetic field planes is the CCR. 
 
It must be mentioned that studies like that of  \citealt{1979ApJ...229..348C} (hereafter CR79) suggest that due to
  propagation effects like adiabatic walking, close to 100 $\%$
  linearly polarized subpulse can emerge at the escape region even if
  the polarization were to be random at the generation point.
  Adiabatic walking is a geometrical effect that
  requires the emission cones to move away from the curved magnetic
  field lines as a result of a rectilinear propagation. The particular
  condition required is that the emission must be at the
  characteristic plasma frequency $\omega_{1}$
  (see Eq. 7 of CR79). MMG14 showed that
  this condition does not hold at the radio emission zones. The radio
  emission is excited and escapes at frequencies much lower than the
  characteristic plasma frequency $\omega_\mathrm{1}$ of the
  outflowing plasma at distances of around 500 km from the
  surface. Hence the origin of highly linearly polarized subpulses
cannot arise due to the mechanism suggested by CR79. 

Considering CCR as the radio emission mechanism, we
estimate the typical value of $\xi$ for the escaping $t$-waves in the
CCR regime ($\omega_\mathrm{CCR} \ll \omega_{1}$),
using the formalism of \citealt{2004ApJ...600..872G} (hereafter GLM04)
as shown in Appendix \ref{tccr}.
$\xi_\mathrm{CCR}$ for $t$-mode in the radiation zone is,
\begin{equation}
    \xi_\mathrm{CCR} \approx \gamma_\mathrm{s}  \gg  1 \label{tmode_ccr}  
\end{equation}
Thus, CCR excites $t$ waves that can freely escape from curved
magnetic field lines with the predominant component of the wave
electric field being perpendicular to the magnetic field planes at the
radio emission zone. This corresponds to the case where $k-B$ planes coincides
with the magnetic field planes.

It must be mentioned that the power in CCR is given in the X-mode perpendicular
to the magnetic field plane and O-mode in the magnetic field
plane. However, the power in O-mode of CCR is Razin suppressed for the
$lt_{2}$ mode and is ducted along the magnetic field lines for
$lt_{1}$ mode until it decays by Landau damping
(\citealt{1986ApJ...302..120A}). It is the power in the X-mode that
escapes as $t$ waves. But, observations show the presence of a
secondary polarization mode. It is identified with the O-mode. The
escape of O-mode from pulsar plasma presents a problem for CCR
regime. MMG14 has suggested that plasma density gradients can aid in
the escape of $lt$ modes. However, such studies that conclusively
shows this effect are yet to be carried out. Secondly, the $t$ mode
and $lt$ modes are linearly polarized. It is difficult to explain the
origin of circular polarization in the CCR regime. Studies like
\citet{1998Ap&SS.262..379L} have suggested a rotation induced
coupling between the linearly polarized eigen modes can result in 
circular polarization. However,
the effect of this coupling in the CCR regime still needs to
be investigated.  

To summarize, while the $t$ mode can freely escape the plasma, any
fictitious mechanism cannot explain the observational features.  Only
$t$ mode excited by CCR can explain certain key observational features
like close to 100\% polarization of sub-pulses that follow the RVM 
and the observed perpendicular orientation of the electric
field with respect to the magnetic field line planes. However, the origin of OPMs and
circular polarization in CCR regime still remains unresolved.

\section{Conclusion}\label{conclusion}

 In this study we obtain constraints on the externally driven
  plasma model of PTS20, that has been proposed to excite superluminal
  $lt_2$ mode as a candidate for explaining pulsar radio emission. The
  model considers the time dependent pair creation process at the
  polar cap as a source of electromagnetic radiation which can
  directly transfer energy to superluminal $lt_2$ mode at a small
  angle of excitation. However, for realistic pulsar parameters (see
  section \ref{obs}) we find that the notion of traveling-wave
  solutions is not suitable for both linear and exponential models of
  pair cascade discharge just above polar cap irrespective of the
  value of plasma multiplicity $\kappa$ and the bulk Lorentz factor
  $\gamma_\mathrm{s}$.  We suggest that any transient energy exchanges
  during discharge can only be channelled into increasing the momentum
  spread of the pair plasma particles. It must be noted here that the
  pair cascade model fails because the associated characteristic
  timescale $\tau$ is very small $\sim$ 100 nanoseconds.  In the
  vacuum gap models, the characteristic timescale $\tau$ is of the
  order of light crossing time of the polar cap radius and hence
  cannot be changed drastically. However, after the discharge when a
  stationary plasma flow is established, fictitious mechanisms
  operating on much longer timescales compared to timescale $\tau$ are
  not ruled out (see e.g. \citealt{2021MNRAS.500.4549M}). As discussed in subsection \ref{disp curves} the
  dispersion relation of $lt_2$ mode at radio frequencies changes with
  plasma multiplicity $\kappa$ and the bulk Lorentz factor
  $\gamma_\mathrm{s}$. The $lt_2$ dispersion relation allows
  travelling-wave solutions for frequencies only above the cut-off
  frequency $\omega_\mathrm{o} \propto \sqrt{\kappa}/
  \gamma^{3/2}_\mathrm{s}$. For a constant $\gamma_\mathrm{s}$, the
  cut-off frequency $\omega_\mathrm{o}$ is higher for higher
  multiplicity and vice versa. As a result, the $lt_2$ dispersion
  relation spans the full broadband radio frequencies closer to the
  neutron star for lower multiplicities and farther away from the
  neutron star for higher multiplicities. This effect is shown in
  Fig. \ref{super_radioband} where for a constant $\gamma_\mathrm{s}
  =100$, the full broad-band radio frequencies are spanned a few tens
  of km and a few hundreds of km above the surface for $\kappa = 10^3$
  and $\kappa = 10^5$ respectively. It must also be noted for a fixed
  $\kappa$, decreasing the bulk Lorentz factor $\gamma_\mathrm{s}$
  increases the cut-off frequency $\omega_\mathrm{o}$ thereby
  increasing the distance where the full radio-band can be
  spanned. Thus, for any fictitious emission mechanism traveling wave
  solutions are possible in regions where the full radio-band can be
  accommodated. However even if the radio-band can be excited at these
  regions, we find that the $lt_2$ modes cannot escape from the plasma
  under homogeneous conditions.

The only mode that can freely escape the plasma is the $t$ mode. However, we
show that for an arbitrary fictitious mechanism the emergent
polarization of the wave is randomly oriented with respect to the
magnetic field line planes.  Averaging of such random polarization inevitably leads to
depolarization of the signal which is in contradiction with
observations where 100\% polarized subpulses are seen that follow the
RVM. However, CCR can excite the plasma modes both parallel and
perpendicular to the field line planes. As a result, the emerging $t$ mode
excited due to CCR can have polarization perpendicular to magnetic
field planes. This aspect is consistent with observations, and to the
best of our knowledge, no other known mechanism(s) satisfies these stringent
constraints.  Hence CCR seems not only viable but a very promising
alternative.

\section*{Acknowledgements}
We thank the anonymous referee for critical comments and suggestions that has greatly improved the quality of the manuscript. We thank Rahul Basu
for discussions and critical comments on the manuscript.
Sk. MR and DM acknowledge the support of the Department of Atomic
Energy, Government of India, under project
no. 12-R\&D-TFR-5.02-0700. DM acknowledges support and funding from
the ‘Indo-French Centre for the Promotion of Advanced Research -
CEFIPRA’ grant IFC/F5904-B/2018.
This work was  supported by the grant 2020/37/B/ST9/02215 of the 
National Science Centre, Poland.

\section*{Data Availability}

No new data were generated or analysed in support of this research.



\bibliographystyle{mnras}
\bibliography{References} 




\appendix

\section{Group velocity of \lowercase{$lt_{2}$} mode}\label{appB}
For simplicity we assume $c = 1$. 
The dispersion relation of the $lt_{2}$ mode are
\begin{equation}
   \epsilon (\omega,k_{\parallel},k_{\perp}) =  \left( \omega^2 - k^2_{\parallel} \right) \left[ 1 -  \frac{\omega^2_\mathrm{p}}{\gamma^3_\mathrm{o}(\omega - k_{\parallel} \mathrm{v}_\mathrm{o} )^2}\right] - k^2_{\perp} = 0 \label{dispersion relation}
\end{equation}

The group velocity of the $lt_{2}$ mode are given by
\begin{equation}
   \vec{v}_\mathrm{gr} =  \frac{\partial \omega}{\partial k_{\perp}} \; \hat{b}_{\perp} + \frac{\partial \omega}{\partial k_{\parallel}} \; \hat{b}_{\parallel} \label{definition}
\end{equation}
where $\hat{b}_{\parallel}$ and $\hat{b}_{\perp}$ are unit vectors parallel and perpendicular to the ambient magnetic field $\vec{B}$.

Taking derivative of equation (\ref{dispersion relation}) with respect to $k_{\perp}$ we get
\begin{equation}
    \frac{\partial \omega}{\partial k_{\perp}} \left[ \Bigg\{ 1 - \frac{\chi^2}{(1 - n_{\parallel} \mathrm{v}_\mathrm{o})^2 }\Bigg\} + \frac{ \chi^2 (1 - n^2_{\parallel}) } {(1 - n_{\parallel} \mathrm{v}_\mathrm{o})^3 }\right] = n_{\perp}  \label{perp derivative}
\end{equation}

Taking derivative of equation (\ref{dispersion relation}) with respect to $k_{\parallel}$ we get 
\begin{equation} \label{parallel derivative}
\begin{split}
   &\ \frac{\partial \omega}{\partial k_{\parallel}} \left[ \Bigg\{ 1 - \frac{\chi^2}{(1 - n_{\parallel} \mathrm{v}_\mathrm{o})^2 }\Bigg\} + \frac{ \chi^2 (1 - n^2_{\parallel}) } {(1 - n_{\parallel} \mathrm{v}_\mathrm{o})^3 }\right] \\ &  = n_{\parallel} \; \Bigg \{ 1 - \frac{n_{\parallel}\chi^2}{(1 - n_{\parallel} \mathrm{v}_\mathrm{o})^2} \Bigg \} + \frac{(1 - n^2_{\parallel}) \chi^2 \mathrm{v}_\mathrm{o}}{ (1 - n_{\parallel} \mathrm{v}_\mathrm{o})^3} 
\end{split}   
\end{equation}

\section{Polarization of \lowercase{$t$} mode for CCR in the radiation zone}\label{tccr}

We refer the reader to the cylindrical co-ordinate system ($r, \theta, z$) GLM04 in fig. 1 in GLM04. The goal is to estimate the orientation of emergent electric field in the radiation zone $r \gg r_\mathrm{o}$ due to $t$ waves generated by CCR bunches located at $r\approx r_\mathrm{o} $.

The wave vector has $\vec{k}$ has the components ($k_{r}, k_{\theta},k_\mathrm{z}$) while the wave electric field $\vec{E}_\mathrm{wave}$ has the components ($E_{r}, E_{\theta},E_\mathrm{z}$). The representation of the wave vector $\vec{k}$ and the wave electric field are connected to the local cartesian system of Fig. \ref{DR_orthogonal} by the transformation
\begin{eqnarray}
   &\ k_{\parallel} = k_{\theta} \\
   &\ k_{\perp} = \sqrt{k^2_\mathrm{z} + k^2_\mathrm{r}} \\ 
   &\ E_{\parallel} = E_{\theta} \\
   &\ E_{\perp} = \sqrt{E^2_\mathrm{z} + E^2_\mathrm{r}}
\end{eqnarray}

GLM04 introduces the dimensionless variables $s$ and $x$ are defined for small angle propagation as 
\begin{eqnarray}
  &\ s = k r_\mathrm{o} \label{s} \\
  &\ x = \left( \frac{r - r_\mathrm{o}}{r}\right) s^{2/3} \label{x} 
\end{eqnarray}
As per this definition $xs^{-2/3} \ll 1$ corresponds to the near-field and $xs^{-2/3} \gg 1$ corresponds to the radiation zone.

For the $t$ mode we have $E_{\parallel} = E_{\theta} = 0$ which gives
 \begin{equation}
     \vec{E} = (E_{r}, 0 , E_\mathrm{z}) 
 \end{equation}

The components $E_{r}$ and $E_{z}$ can be estimated from the vector potential $\vec{A}(A_r,A_{\theta},0)$ from (see Eq. 4 and 6 of GLM04)
\begin{eqnarray}
  &\ E_{r} = i k_\mathrm{z} A_{\rho}  \label{Ez1}\\
  &\ E_{z} = - k \left( i A_{\theta} + s^{-1/3}   \frac{\partial A_{\rho}}{\partial x} \right)  \label{Er1}
\end{eqnarray}
where the $t$ mode dispersion relation  $\omega/c = k$ has been used for simplification.

In the CCR regime, the following relationship holds among the vector potential (see Eq. 24 of GLM04)
\begin{equation}
A_{\theta} = - \frac{i s^{1/3}}{2 x} \frac{\partial A_{r}}{\partial x}  \label{Atheta1} 
\end{equation}
In the radiation zone the vector potential $A_{r}$ has the asymptotic solution (see Eq. 26 of GLM04)
\begin{equation}
    A_{r} = \exp\left( i \frac{2 \sqrt{2}}{3} x^{3/2}\right) \label{Ar}
\end{equation}

Using equation (\ref{Ar}) in equation (\ref{Atheta1}) we get,
\begin{equation}
    A_{\theta} = \frac{s^{1/3}}{\sqrt{2x}} \exp\left( i \frac{2 \sqrt{2}}{3} x^{3/2}\right)  \label{Atheta2}
\end{equation}

Using equation (\ref{Ar}) and equation (\ref{Atheta1}) in equation (\ref{Ez1}) and equation (\ref{Er1}) gives us
\begin{eqnarray}
    &\ E_\mathrm{z} = -  i k \left(  \frac{s^{1/3} + 2 x s^{-1/3}}{\sqrt{2x}} \right)  \; \exp \left( i \frac{2 \sqrt{2}}{3} x^{3/2} \right) \\
    &\ E_{r} = i k_\mathrm{z}  \; \exp \left( i \frac{2 \sqrt{2}}{3} x^{3/2} \right) 
 \end{eqnarray}

In the radiation zone we have $xs^{-2/3} \gg 1$ which reduces the equation to  
\begin{eqnarray}
    &\ E_\mathrm{z} = -  i k \sqrt{2x s^{-2/3}}  \; \exp \left( i \frac{2 \sqrt{2}}{3} x^{3/2} \right) \label{Ezfinal} \\
    &\ E_{r} = i k_\mathrm{z}  \; \exp \left( i \frac{2 \sqrt{2}}{3} x^{3/2} \right) \label{Erfinal} 
 \end{eqnarray}

For CCR we have $k/k_{z} \approx \gamma_\mathrm{s}$ and $s = \gamma^3_\mathrm{s}$. We define a dimensionless quantity  $\xi_\mathrm{CCR}$ in the radiation zone from equation (\ref{Ezfinal}) and (\ref{Erfinal}) as
  \begin{equation}
     \xi_\mathrm{CCR} = \left| \frac{E_\mathrm{z}}{E_{r}} \right| \approx  \sqrt{x} 
 \end{equation}
 In the radiation zone we have $x s^{-2/3} \gg 1$ which translates to the condition $x \gg s^{2/3}$. Using $s = \gamma^3_\mathrm{s}$ we have $x \gg \gamma^2_\mathrm{s}$. This condition reduces the equation above to the form
  \begin{equation}
     \xi_\mathrm{CCR} = \left| \frac{E_\mathrm{z}}{E_{r}} \right|_\mathrm{radiation-zone}  \gg \gamma_\mathrm{s} \gg 1 \label{xiccr}
 \end{equation}

 Thus, in the radiation zone the  emergent radio waves due to  $t$ waves excited by CCR emerges predominantly polarized perpendicular to the magnetic field plane. It must be noted that the Eq. 18 of GLM04 represents the wave electric field in the near-field, i.e. $\xi_\mathrm{CCR, near-field} = - s^{1/3} k/(k_z \sqrt{2 x}) $.
 
 As a final check of consistency, we investigate whether the waves at the radiation zone can freely escape the plasma. To do so it is necessary that the waves in the radiation zone retain a transverse character while satisfying the small wavelength approximation (see Eq. B4 of GLM04) expressed as 
\begin{equation}
    k^2_{r} \approx  2 \frac{s^{4/3} x}{r^2_\mathrm{o}} \label{kexpr1}
\end{equation}

The transverse character of the emergent waves require ($\vec{k} \cdot \vec{E} = 0$)
\begin{equation}
      k_{r} = - \left( \frac{E_\mathrm{z}}{E_\mathrm{r}} \right) k_\mathrm{z} \label{kexpr2}
\end{equation}
 By substituting equation (\ref{Erfinal}) and equation (\ref{Ezfinal})  in equation (\ref{kexpr2}) we get,
 \begin{equation}
      k^2_{r} = k^2_\mathrm{z} \left(\frac{E_\mathrm{z}}{E_{r}}\right)^2 = k^2 \; 2x s^{-2/3} = \frac{s^2}{r^2_\mathrm{o}} \; 2 x s^{-2/3}  \label{kexpr3}
 \end{equation}
 
 Thus, the equality of the expression can be seen from comparing equation (\ref{kexpr1}) and equation (\ref{kexpr2}).
 
 To summarize, $t$ mode excited by CCR bunches can escape as freely propagating transverse electromagnetic radiation in the radiation zone such that the predominant component of the wave electric field is perpendicular to the magnetic field plane. 

\section{Notations and symbols used throughout the text}

\begin{itemize}
    \item $P$: Period of pulsar.
    \item $B_\mathrm{d}$: Strength of the global magnetic field of the pulsar.
    \item $B_\mathrm{s}$: Strength of the non-dipolar surface magnetic field.
    \item $b$: Ratio of $B_\mathrm{s}$ to $B_\mathrm{d}$ at the neutron star surface. 
    \item $\rho_\mathrm{c}$: Radius of curvature of the magnetic field lines.
    \item $n_\mathrm{GJ}$: Goldreich-Julian number density. 
    \item $\kappa$: Multiplicity of pair plasma.
    \item $n_\mathrm{s}$: Number density of pair plasma.
    \item $\gamma_\mathrm{s}$: Bulk lorentz factor of the pair plasma.  
    \item $\omega$: The angular frequency of a plasma wave.
    \item $k$: The wave number of a plasma wave
    \item $\Vec{k}$: wave vector of the plasma waves.
    \item $\Vec{B}$: The local ambient pulsar magnetic field. 
    \item $k-B$ plane: The plane containing $\vec{k}$ and $\vec{B}$.
    \item $\vartheta$:The angle between $\vec{k}$ and $\vec{B}$ as shown in Fig. \ref{Ortho_coord}. Referred to as angle of excitation.
    \item $t$-mode: The purely transverse mode perpendicular to the $k-B$ plane.
    \item ordinary $lt$ mode: The longitudinal-transverse plasma mode confined to the $k-B$ plane.
    \item Langmuir $L$ mode: The purely longitudinal and electrostatic ordinary mode for $\vartheta = 0$.
    \item $TE$-mode: The purely transverse $lt$ mode for $\vartheta = \pi/2$.
    \item $lt_{1}$ waves: The subluminal branch of the $lt$ mode for $\vartheta > 0$.
    \item $lt_{2}$ waves: The superluminal branch of the ordinary mode for $\vartheta > 0$.
    \item $\omega_\mathrm{p}$: The plasma frequency of the pair plasma as defined in equation (\ref{plasma_freq}).
    \item $\omega_\mathrm{o}$: The cut-off frequency of Langmuir mode and $lt_{2}$ waves defined in equation (\ref{cut-off}).
    \item $\omega_{1}$: The characteristic frequency of the plasma as defined in equation (\ref{characteristic}).
    \item $\vec{E}_\mathrm{wave}:$ The wave electric field of the $lt_{2}$ waves. 
    \item $\vec{E}_{2}$: The wave electric field of $t$ wave perpendicular to the $k-B$ plane.
    \item $E_{1}$: Projection of $\vec{E}_\mathrm{wave}$ perpendicular to $\vec{k}$.
    \item $E_{3}$: Projection of $\vec{E}_\mathrm{wave}$ parallel to $\vec{k}$.
    \item $E_{\perp}$: Projection of $\vec{E}_\mathrm{wave}$ perpendicular to $\vec{B}$.
    \item $E_{\parallel}$: Projection of $\vec{E}_\mathrm{wave}$ parallel to $\vec{B}$.
    \item $\Theta $: The angle between wave electric field $\vec{E}_\mathrm{wave}$ and $\vec{k}$ as defined in equation (\ref{tantheta}). 
    \item  $k_{\perp}$: Projection of $\vec{k}$ perpendicular to $\vec{B}$.
    \item $k_{\parallel}$: Projection of $\vec{E}_\mathrm{wave}$ parallel to $\vec{B}$.
    \item $n = k c/\omega$: The refractive index of plasma for a plasma wave of frequency $\omega$ and wave number $k$.
    \item $n_{\parallel} =  n \cos \vartheta $: parallel refractive index of the plasma.
    \item $n_{\perp} = n \sin \vartheta$: perpendicular refractive index of the plasma.
    \item $\vec{v}_\mathrm{gr}$: The group velocity of $lt_{2}$ waves.
    \item $\vartheta_\mathrm{gr} $: Angle between the ambient magnetic field $\vec{B}$ and group velocity $\vec{v}_\mathrm{gr}$ of the $lt_2$ mode as defined in equation (\ref{grp_vel_angle}).
    \item $\tau:$ The characteristic timescale associated with pair cascade process in the polar cap as defined in equation (\ref{tau}).
    \item $t_\mathrm{max}$: The maximum timescale of discharge as defined in equation (\ref{tmax}).
    \item $\vec{\Omega}$: Rotation vector of the pulsar.
    \item Magnetic field plane: The plane containing $\vec{\Omega}$ and $\vec{B}$ as shown in Fig. \ref{Magnetic_field_plane}.
    \item $\phi$: The angle between $\vec{k}$ and the rotation axis $\vec{\Omega}$ as shown in Fig. \ref{Magnetic_field_plane}.
    \item $\delta$: The angle between the projection of the wave vector on the plane perpendicular to the magnetic field plane and the ambient magnetic field $\vec{B}$ as shown in Fig. \ref{Magnetic_field_plane}.
    \item CCR: Coherent Curvature Radiation
    \item $\xi_\mathrm{CCR}$: The ratio of $t$ mode electric field parallel and perpendicular to the magnetic field plane as defined in equation (\ref{tmode_ccr}).
\end{itemize}


\bsp	
\label{lastpage}
\end{document}